\documentclass[a4paper,fleqn,usenatbib]{mnras}

\usepackage[T1]{fontenc}
\usepackage{ae,aecompl}

\usepackage{graphicx}	% Including figure files
\usepackage{amsmath}	% Advanced maths commands
\usepackage{amssymb}	% Extra maths symbols

\usepackage{hyperref}
\usepackage{color}
\usepackage{soul}
\usepackage{epstopdf}
\usepackage{booktabs,caption}
\usepackage[flushleft]{threeparttable}
\usepackage{blindtext}

\title[Evolution of the PPDs]{Time-dependent evolution of the protoplanetary discs with magnetic winds}
\author[M. Shadmehri \& S. M. Ghoreyshi]{
Mohsen Shadmehri$^{1,2}$\thanks{E-mail: m.shadmehri@gu.ac.ir} and Sayyedeh Masoumeh Ghoreyshi$^{1}$
\\
$^{1}$  Department of Physics, Faculty of Sciences, Golestan University, Gorgan 49138-15739, Iran\\
$^{2}$Research Institute for Astronomy and Astrophysics of Maragha (RIAAM), Maragha, P.O. Box: 55134-441, Iran\\
}

\date{Accepted XXX. Received YYY; in original form ZZZ}

\pubyear{2017}

\begin{document}
\label{firstpage}
\pagerange{\pageref{firstpage}--\pageref{lastpage}}
\maketitle

\begin{abstract}
We study the evolution of the protoplanetary discs (PPDs) in the presence of magnetically driven winds with the stress relations motivated by the non-ideal MHD disc simulations. Contribution of the magnetic winds in the angular momentum removal and mass loss are described using these relations which are quantified in terms of the plasma parameter. Evolution of the essential disc quantities including the surface density, accretion rate and wind mass loss rate are studied for a wide range of the model parameters. Two distinct phases of the disc evolution are found irrespective of the adopted input parameters. While at the early phase of the disc evolution, global disc quantities such as its total mass and magnetic flux undergo non-significant reductions, their rapid declines are found in the second phase of evolution. Duration of each phase, however, depends upon the model parameters including magnetic wind strength. Our model predicts that contributions of the magnetic winds in the disc evolution are significant during the second phase. We then calculated locus of points in the plane of the accretion rate and total disc mass corresponding to an ensemble of evolving PPDs. Our theoretical isochrone tracks exhibit reasonable fits to the observed PPDs in star forming regions Lupus and $\sigma$-Orion. 

\end{abstract}


%%%%%%%%%%%%%%%%%%%%%%%%%%%%%%%%%%%%%%%%%%%%%%%%%%

%%%%%%%%%%%%%%%%% BODY OF PAPER %%%%%%%%%%%%%%%%%%

\section{Introduction}

Understanding angular momentum transport mechanisms in the protoplanetary discs (PPDs) is an important issue because of its starring role  in the disc lifetime and the outcome of planet formation.  In the conventional paradigm of the disc accretion, the  magnetorotational instability \citep[MRI;][]{Balbus1991} is suggested to be efficient in the well-ionized regions, whereas in the outer parts of the PPDs where are not sufficiently ionized, the gravitational instability is probably a dominant mechanism \citep[see][for a review]{Krat2016}. There are, however, alternative processes based on the purely hydrodynamics instabilities, including convective overstability \citep[e.g.,][]{Klahr2014}, vertical shear instability \citep[e.g.,][]{Nelson2013} and the zombie vortex instability \citep[e.g.,][]{Marcus2015}. It is therefore an enduring challenge to determine which physical process is responsible for the PPDs angular momentum transport. 

In the recent years, on the other hand, MHD disc simulations have also shown that wind launching is directly related to the MRI-driven  turbulence   in the presence of the non-ideal MHD effects \citep{Suzuki2010,Bai2013,Lesur2013,Bet2017}. This important finding then implies that disc accretion may occur via angular momentum removal by winds instead of MRI which redistributes angular momentum within the disc. Robust diagnostics thereby are needed to discriminate between a standard viscous disc model  \citep[][ hereafter SS73]{Shakura1973}  and the magnetic wind-driven accretion scenario as a dominant mechanism of the accretion in the PPDs.

The viscous and the wind-driven accretion disc models predict distinct features which can be reconciled with the observational constraints on the PPDs structure. Measurements of the disc radii \citep{Najita2018}, the turbulent strength \citep{Wang2017,Rafikov2017} and exploring "disc isochrones" in the accretion rate-disc mass plane \citep[][hereafter LSMT17]{Lodato17} provide useful insights into mechanisms that derive the accretion within the PPDs. In the presence of the viscous spreading mechanisms, \cite{Najita2018} argued that disc size of Class II sources could be larger than their earlier evolutionary, Class I gas discs. They found that most Taurus Class II discs are larger compared to Class I discs. But the analysis of \cite{Najita2018} did not provide a robust signature of MRI because of the neglected processes, such as photoevaporation, that may affect disc size measurements. The idea behind this study, however, is interesting to be explored further using a much larger number of sources and more sophisticated models for measuring disc radii. 

\cite{Wang2017} argued that the accretion speed in a few well-studied discs is transsonic and it is a natural consequence of the accretion driven by the magnetized winds because MRI driven turbulence is not strong enough to account for the high levels of the observed accretion. Recent observations of the PPDs residing in Chamaeleon I suggest that magnetic winds may play a role in the angular momentum transport \citep{Mulders2017}. In another study by \cite{Rafikov2017}, he used a sample of 26 PPDs resolved by ALMA to estimate the dimensionless viscosity parameter $\alpha$ if the standard viscous disc model (SS73) is implemented to describe these systems. The inferred value spans a broad range from $10^{-4}$ to 0.04 and the central accretion rate is found to be decoupled from the global accretion rate probably owing to magnetohydrodynamic winds \citep{Rafikov2017}. There are, however, studies that suggest the observed accretion rates of the PPDs can be explained with a low viscosity as a result of hydrodynamics \citep{Hartmann2018}. 

Despite these recent achievements, none of them provides a verified observational signature because of difficulties in detecting disc winds and spatially resolve them and theoretical uncertainties in the disc models with winds. Therefore, it is an important issue to explore the evolution of the PPDs in the presence of magnetically driven winds. 
In the early models of the magnetic winds, the primary approach was to investigate this phenomenon independent of underlying  accretion disc \citep[e.g.,][]{Blandford1982,Konigl1989,Ferreira1995}. The magnetically driven winds are able to extract the mass, angular momentum and energy from the disc, however, their rates are not well-constrained owing to their relations with the disc structure itself and the associated angular momentum transport and removal mechanisms. Numerical MHD disc simulations are generally used to address wind launching conditions and to quantify rates of the  mass loss and the angular momentum removal \citep[e.g.,][]{Lesur2013,Simon2013a,Fromang2013,Bet2017}.  

Recent disc models incorporated wind effects motivated by recent MHD simulations or physical arguments within the framework of the standard accretion disc model \citep{Suzuki2010,Armitage2013,Suzuki2016,Bai2016}. The simplicity of this approach is an advantage to investigate disc evolution for a broad range of the input parameter space. \cite{Suzuki2010}  presented a simple 1D model of PPD with winds inspired by their MHD disc simulations. They did not consider wind torque, however, wind mass loss was included in their study. \cite{Suzuki2016} presented a mode generalized model with the stress components of the disc turbulence and wind launching and mass loss effect. They found a wide variety of the disc surface density profile even with a positive slope  depending on the adopted model parameters. The disc surface density profile with a positive slope may suppress the type I migration of protoplanets \citep{Ogihara2015,Ogihara2018}. \cite{Suzuki2016}  did not study the evolution of important disc quantities such as size, mass and accretion rate to be confronted with the observed PPDs. \cite{Armitage2013} also studied disc evolution with winds, however, they did not include wind mass loss and the implemented relations for the disc turbulence and wind were parameterized in terms of the plasma parameter using MHD disc simulations \citep{Simon2013a}. Subject to the steady state approximation, this model leads to the fully analytical solutions \citep{khajenabi2018} which were used in other research directions such as explaining ring like structures in HL Tau \citep{Hasegawa2017,khajenabi2018} or pebble production in the PPDs with the magnetic winds \citep{shadmehri2018}. But the aforementioned  studies neither provide detailed physical insights about wind kinematics nor reconcile the obtained trends with the observed features of the PPDs. 

We describe our disc model with magnetic winds and the implemented assumptions in Section 2. Disc evolution properties, including isochrone tracks, are explored in Section 3. In Section 4, we provide astrophysical implications of our model by comparing our theoretical isochrone tracks with the observed PPDs in Lupus and $\sigma$-Orion. Finally, we summarize and conclude in Section 5.

\section{Basic Equations}
Our model is constructed based on the standard accretion disc model (SS73) and the magnetically driven winds are incorporated via their roles in the mass loss and angular momentum removal. The disc is assumed to rotate with a Keplerian profile, i.e., $\Omega=\sqrt{GM_{\star}/r^3}$ where $M_{\star}$ is the star mass and $r$ denotes the radial distance. This assumption holds true as long as contribution of the magnetic force in the radial direction is negligible and the disc remains cold enough. We also do not include disc compression in the vertical direction due to the magnetic force. Therefore, disc hydrostatic equilibrium in the vertical direction implies that $H=\sqrt{2} c_{\rm s}/\Omega$,  where $H$ and $c_{\rm s}$ are the disc scale height and the sound speed. A similar model has been implemented in prior models \citep[e.g.,][]{Armitage2013,Suzuki2016,Bai2016}, but either all these wind effects have not been included simultaneously or even in the presence of the mass loss and angular momentum removal the associated coefficients were set somehow uncorrelated. 

We instead consider a thin accretion disc that undergoes viscous evolution with mass loss and angular momentum removal owing to the magnetically driven winds. Thus, the following partial differential equation is obtained \citep{Suzuki2016}:
\begin{align}\label{eq:main1}
 \nonumber \frac{\partial \Sigma}{\partial t}=\frac{1}{r}\frac{\partial}{\partial
 r}\left [ \frac{2}{r\Omega}\left ( \frac{\partial}{\partial r} \left (r^2\Sigma{c_{\rm s}}^2 W_{r\phi}\right )+r^2{(\rho c_{\rm s}^2)_{\rm mid}} W_{z\phi} \right ) \right ] \\ 
 -C_{\rm w} (\rho c_{\rm s})_{\rm mid},   
 \end{align}
where $\Sigma$ stands for the disc surface density and the midplane density is  $\rho_{\rm mid}=\Sigma/(\sqrt{\pi}H$) and the sound speed is defined as $c_{\rm s}=\left ( k_{\rm B}T/\mu m_{\rm H} \right )^{1/2}$. Here, $T$ and $\mu$ are the temperature and mean molecular weight and $k_{\rm B}$ and $m_{\rm H}$ stand for the Boltzmann constant and the hydrogen mass, respectively. We suppose that the mean molecular weight is $\mu =2.1$ \citep[e.g.,][]{shadmehri2018}. In equation (\ref{eq:main1}), $W_{r\phi}$ and $W_{z\phi}$ are the normalized accretion stresses owing to the MHD disc turbulence and the magnetic wind, respectively. Mass removal by wind is parameterized by the last term where a coefficient, i.e., $C_{\rm w}$, is introduced to control wind mass loss rate \citep{Suzuki2016}. \cite{Armitage2013} used an equation similar to equation (\ref{eq:main1}) above, but in the absence of the wind mass loss, i.e., $C_{\rm w}=0$. 

The normalized accretion stresses are defined as \citep{Simon2013b,Suzuki2016,Hasegawa2017}
\begin{equation}
W_{r\phi}=\frac{\int_{-H_{\rm w}}^{+H_{\rm w}} \big(\rho v_r\delta v_{\phi}-B_rB_{\phi}/4\pi \big) dz}{\Sigma c_{\rm s}^2},
\end{equation}
\begin{equation}
W_{z\phi}=2|W_{z\phi}|_{\rm bw}=\frac{\big(\rho v_z\delta v_{\phi}-B_z B_{\phi}/4\pi \big)_{z=-H_{\rm w}}^{z=+H_{\rm w}}}{(\rho c_{\rm s}^2)_{\rm mid}},
\end{equation}
Note that $W_{z\phi}$ stands for the normalized angular momentum flux by the winds from the top and bottom surface of the disc. It is assumed that wind launching regions are disc surfaces  at the distance $z=\pm H_{\rm w}$ from the disc midplane. The exact value of $H_{\rm w}$ was discussed in \cite{Simon2013b}, however, we assume that $H_{\rm w} \simeq 2 H$ \citep{Hasegawa2017}. The quantity $|W_{z\phi}|_{\rm bw}$ represents the value of $W_{z\phi}$ at the base of the wind, i.e., $z=\pm H_{\rm w}$. Subject to the even-$z$ symmetry for the wind, we have $W_{z\phi}=2|W_{z\phi}|_{\rm bw}$ \citep{Simon2013b}.

Using MHD disc simulations, the above mentioned stress tensor components can be written in terms of the disc midplane plasma parameter $\beta_0$ \citep[]{Armitage2013,Simon2013b,Hasegawa2017}:
\begin{equation}\label{eq:wrphi}
\log W_{r\phi}=-2.2+0.5\tan^{-1}\big(\frac{4.4-\log \beta_0}{0.5} \big) ,
\end{equation}
\begin{equation}\label{eq:wzphi}
\log |W_{z\phi}|_{\rm bw}=1.25-\log \beta_0 ,
\end{equation}
where $\beta_0$ is defined as the ratio of the gas pressure to the magnetic pressure, i.e., $\beta_0 = 8P_{\rm mid}/{B_z}^2$. Note that $P_{\rm mid}=\rho_{\rm mid} {c_{\rm s}}^2$ and $B_z$ are the gas pressure at the disc midplane and the vertical component of the magnetic field, respectively. The equations (\ref{eq:wrphi}) and (\ref{eq:wzphi}) are valid if the parameter $\beta_0$ is taken between $10^3$ (strong wind) and $10^5$ (weak wind) and beyond this interval further MHD disc simulations are needed to find appropriate relations for $W_{r\phi}$ and $W_{z\phi}$. In the case of strong wind, the above relations imply that the normalized stress due to wind, $W_{z\phi}$, is larger than the normalized stress owing to MHD disc turbulence. In this wind-dominated case, efficient angular momentum removal occurs instead of redistributing the angular momentum and thereby the outward disc expansion is suppressed effectively. We also note that the quantity $|W_{z\phi}|$  in \cite{Armitage2013} apparently represents our  $|W_{z\phi}|_{\rm bw}$, however, \cite{khajenabi2018} utilized the fitted relation (\ref{eq:wzphi}) above for $W_{z\phi}$. \cite{Suzuki2016} also introduced quantities $\bar{\alpha}_{r\phi}$ and $\bar{\alpha}_{z\phi}$ corresponding to our $W_{r\phi}$ and $W_{z\phi}$, but they adopted values of these quantities  based on the disc energy budget rather than plasma parameter.

During the evolution of a disc, the midplane parameter $\beta_0$ may undergo spatial and temporal variations which can not be evaluated without imposing a further physical condition. Theoretical studies to constrain magnetic flux transport through a disc do not address this issue properly owing to disregarding non-ideal MHD effects or potential role of the wind-driven accretion \citep{Lubow1994,Okuzumi2014,Guilet2014}. Nevertheless, \cite{Armitage2013}  simply  assumed that the spatial dependence of the midplane $\beta_0$ remains uniform, whereas its temporal evolution is subject to the magnetic flux conservation.  \cite{Bai2016}, however, studied disc evolution by following two extreme scenarios. The first scenario, as in \cite{Armitage2013}, relies on the magnetic flux conservation, while in the second scenario, the magnetic flux is assumed to vary in proportion to the total disc mass. We instead explore disc evolution with a fixed midplane parameter $\beta_0$ and the corresponding magnetic flux and its possible relation with the total disc mass are examined. 

For simplicity, we do not consider the energy balance through the disc including possible heating and cooling mechanisms. Instead, the temperature profile of the disc is given as a power-law function of the radial distance, i.e.,
\begin{equation}\label{eq:temp}
T=T_{0} (\frac{r}{r_0})^{-q},
\end{equation}
where $T_0 =300$ K, $r_0 = 1$ au and the temperature exponent is $0<q\leq1$ \citep[]{D'Alessio2001,Okuzumi2016}. In a more realistic model, however, an energy equation is required. But our simplified approach is adequate to illustrate wind effects. Although we adopt $q=1/2$ as  a standard value in our study, the role of this parameter in the disc evolution is explored in subsection \ref{sec:q}  by considering different values of $q$ within the allowed range. 

The mass accretion rate, $\dot{M}_{\rm acc}$, is also an important quantity that can be calculated using the following relation \citep{Hasegawa2017}:
\begin{equation}\label{eq:acc}
\dot{M}_{\rm acc}= \frac{4\pi}{r\Omega}\left [ \frac{\partial}{\partial r} \left (r^2\Sigma {c_{\rm s}}^2 W_{r\phi} \right )+r^2(\rho {c_{\rm s}}^2 )_{\rm mid}W_{z\phi} \right ].   
\end{equation}
The first term represents the mass accretion rate associated with the disc turbulence and the second term corresponds to the mass accretion rate driven by the wind. In the standard disc model (SS73), the contribution of the magnetic winds in the accretion rate is neglected and only viscous mechanism plays a dominant role in the mass accretion rate and angular momentum transport through the disc. In our model, contributions of both viscous process and the magnetic winds are incorporated and their relative importance depends on the magnetic field strength.

Disc dispersal time scale is determined  using the rate of mass removal from the disc owing to the accretion onto the central star and the wind mass loss. It is thereby important to calculate the cumulative wind mass loss rate enclosed within the disc size, $\dot{M}_{\rm wind}$, using the following equation: 
\begin{equation}\label{eq:Mwind}
\dot{M}_{\rm wind}=2\pi\int_{r_{\rm in}}^{r_{\rm out}}r C_{\rm w} (\rho c_{\rm s})_{\rm mid} dr,
\end{equation}
where $r_{\rm in}$ and $r_{\rm out}$ denote the inner and outer edges of the disc. Obviously, if we set $C_{\rm w}=0$, then wind mass removal is not allowed and the disc undergoes viscous evolution and the magnetic winds contribute only to the angular momentum removal.

\section{Analysis of the disc evolution}
\subsection{Numerical Method and Boundary and Initial Conditions}
Now, we can investigate the evolution of a disc with the magnetically driven winds by solving equation (\ref{eq:main1}) numerically subject to the appropriate boundary conditions with a given initial surface density profile. It is therefore convenient to rewrite our main equation (\ref{eq:main1}) into a dimensionless form: 
\begin{equation}\label{eq:main2}
\frac{\partial y}{\partial \tau}= \frac{1}{x} \frac{\partial}{\partial x} \Big[ \sqrt{x} \frac{\partial}{\partial x}(\xi_r x^{2-q} y) +\xi_{z} x^{(2-q)/2} y \Big]-\xi_{\rm w} x^{-3/2} y,   
\end{equation}
where the dimensionless variables are
\begin{equation}
y=\frac{\Sigma}{\Sigma_0}, x=\frac{r}{r_0}, \tau=\frac{t}{t_0}.
\end{equation}
where $\Sigma_0=1700~{\rm g \hspace{1mm}cm^2}$, $t_0=1/\Omega_0$ and $\Omega_0=(GM_{\odot}/{r_0}^3)^{1/2}=2\times10^{-7}{\rm s}^{-1}$. Note that for obtaining equation (\ref{eq:main2}) above, we used the power-law temperature relation, i.e., equation (\ref{eq:temp}). In the equation (\ref{eq:main2}), the dimensionless parameters $\xi_r$, $\xi_z$, and $\xi_{\rm w}$ are defined as
\begin{equation}
\xi_r=\frac{1}{\sqrt{m_{\star}}}\big(\frac{H_0}{r_0} \big)^2 W_{r\phi},
\end{equation}
\begin{equation}
\xi_z=\frac{1}{\sqrt{\pi}}\big( \frac{H_0}{r_0}\big)W_{z\phi},
\end{equation}
\begin{equation}
\xi_{\rm w}=\sqrt{\frac{m_{\star}}{2\pi}}C_{\rm w},
\end{equation}
where $m_{\star}=M_{\star}/M_{\odot}$. Here, we have $H_0 =\sqrt{2} c_{s0}/\Omega_0\simeq0.051$ au and $c_{s0}=\sqrt{k_{\rm B}T_0 /\mu m_{\rm H}}\simeq 1086.5$ m/s. Thus, we obtain $\xi_r \simeq (0.0026/\sqrt{m_\star}) W_{r\phi}$ and $\xi_z \simeq  0.0288 W_{z\phi}$.  Since  rate of the wind mass loss is not well-constrained by the observations or theoretical models, we parameterize the associated coefficient in terms of the angular momentum removal, i.e., $\xi_{\rm w}=\psi \xi_z$ where the wind parameter $\psi$ is adopted within a range of 0.001 to 0.5. For $\psi <0.001$, we found that wind mass loss rate is not high enough to affect disc evolution, whereas for $\psi>0.5$ the wind mass loss rate becomes so high  that disc is depleted during a very short period that is not supported by the current estimates of the PPDs lifetimes. 

We solve our main equation (\ref{eq:main2}) numerically using an implicit finite difference method with a logarithmically spaced radial grid containing 258 cells. The inner edge is $r_{\rm in}=0.05$ au, whereas the outer boundary is assumed to be $r_{\rm out}=20000$ au. We adopt a very large outer boundary to ensure the accuracy of our results. For the boundary conditions, we set the surface density at the inner edge equal to zero and the radial velocity at the outer edge is assumed to be zero. The initial gas surface density is \citep[e.g.,][]{Alexander2012,Mshadmehri2018}
\begin{equation}
\Sigma (r,0)=\frac{M_{\rm d0}}{2\pi \left [ \exp(-\frac{r_{\rm in}}{r_{\rm d0}})-e^{-1} \right ] r_{\rm d0} r} \exp (-r/r_{\rm d0}),
\end{equation}
where we suppose that $r_{\rm d0}=30$ au. The star mass and the initial disc mass are $M_{\star}=1M_{\odot}$ and $M_{\rm d0}=0.1 M_{\star}$, respectively, unless otherwise is stated.

\subsection{Surface Density Evolution}

\begin{figure}
\includegraphics[scale=1.0]{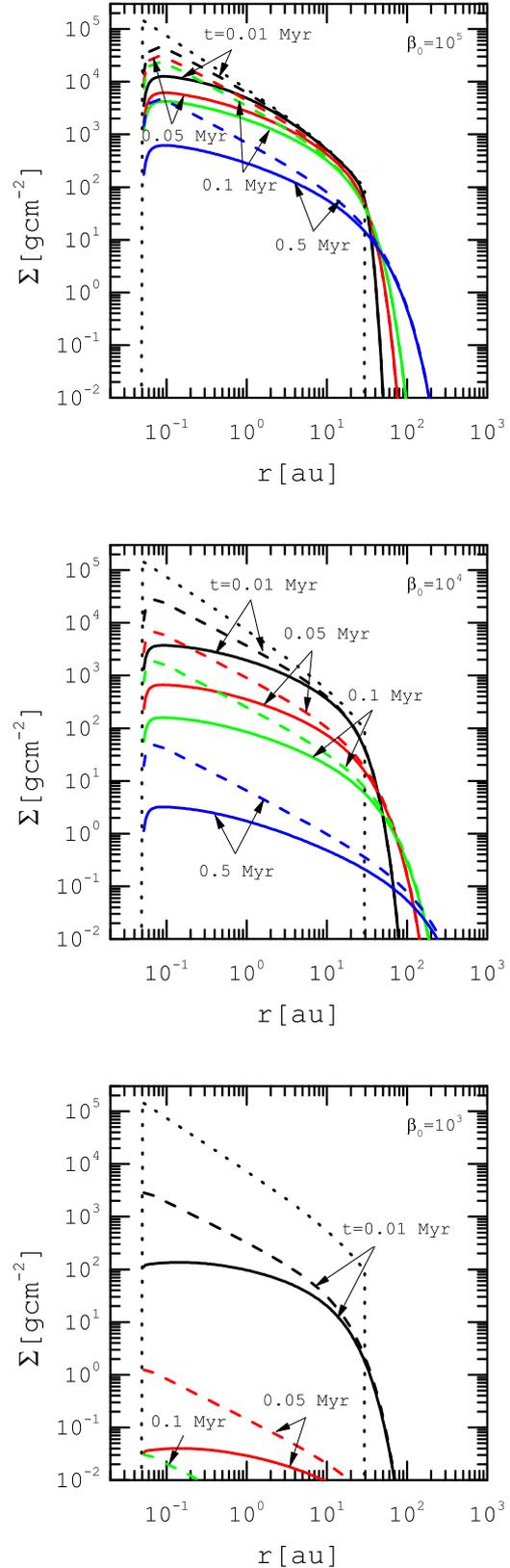}
\caption{Profiles of the disc surface density for different midplane plasma parameter: $\beta_0 =10^5$ (top), $10^4$ (middle) and $10^3$ (bottom).  The initial disc mass is $M_{\rm d0}=0.1 M_{\odot}$ and the temperature exponent is $q=1/2$.  The solid and the dashed curves correspond to $\psi=0.5$ and $\psi=0.001$, respectively.}\label{fig:f1}
\end{figure}

To investigate the role of the magnetic field strength, we first fix all the model parameters at their fiducial values except the midplane parameter $\beta_0$ that is allowed to vary within the permitted interval. Figure \ref{fig:f1} displays surface density evolution at different times, as labeled, and for different levels of the magnetic field strength: $\beta_0 = 10^5$ (top), $10^4$ (middle), and $10^3$ (bottom). The initial surface density distribution is shown by a dotted curve. We also consider different values of the mass removal parameter $\psi$. Solid curves correspond to a disc evolution with a high rate of the wind mass loss, i.e., $\psi =0.5$, whereas the dashed curves display surface density evolution with a lower mass loss rate, i.e., $\psi =0.001$.

The disc surface density gradually decreases with time owing to the viscous and winds effects. But this reduction strongly depends on the magnetic field strength. While in a case with a weak wind, i.e., $\beta_0 =10^5$, and a small mass-loss rate, i.e., $\psi=0.001$, reduction of the surface density at the inner region is about factor of 10  during half a million years, this reduction factor is considerably enhanced to about 600 for $\beta_0 = 10^4$ during the same time period. Note that for this comparison we are considering disc structure at the times  $t=0.01$
and 0.5 Myr, because by the time $0.01$ Myr the disc undergoes viscous evolution and its structure is not the same as the imposed initial state. The reduction of the surface density is so efficient in the strong wind case with $\beta_0 =10^3$ that the disc effectively becomes gas depleted in less than $10^5$ yr. A similar trend is found when the disc is allowed to lose a significant fraction of its mass by the winds. If we compare the evolution of the discs with a given magnetic field strength and different wind mass loss parameter, i.e., $\psi=0.001$ and 0.5, we see that the surface density reduction owing to  an efficient mass removal starts from the inner disc region and gradually proceeds to the outer parts with a rate which depends on the magnetic field strength. While in the weak wind case, the wind mass removal only affects the inner disc parts, as the wind becomes stronger, the entire disc is affected by the wind mass loss.

\subsection{Disc Mass and Magnetic Flux Distribution}

\begin{figure}
\includegraphics[scale=1.0]{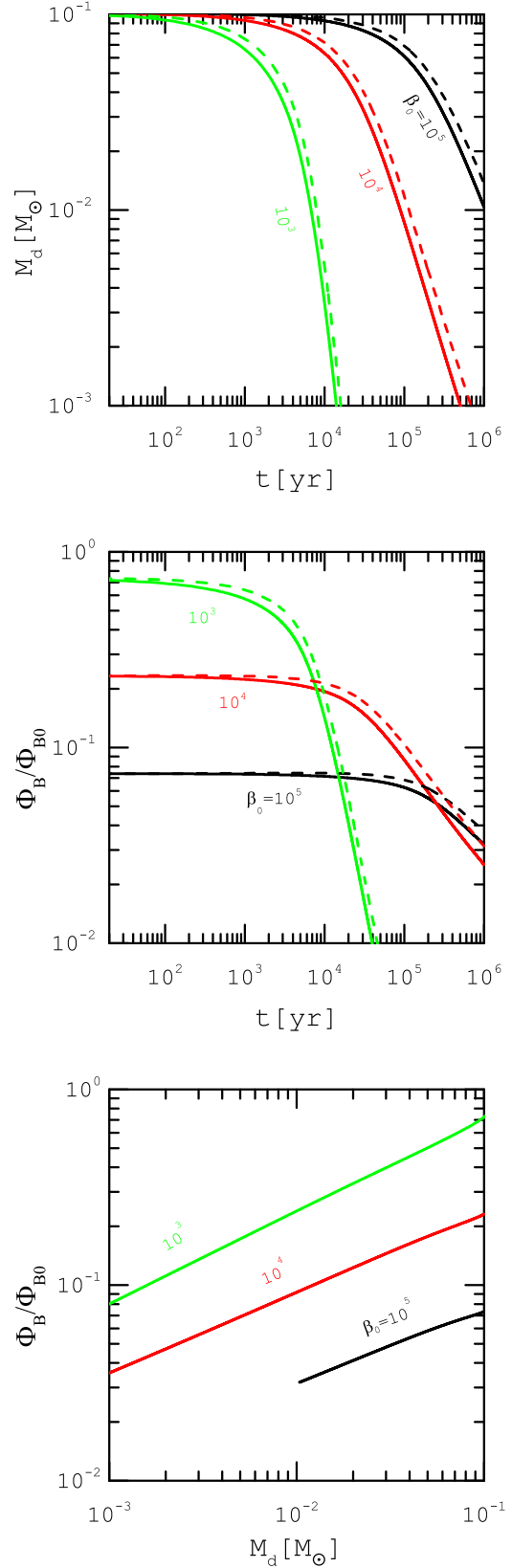}
\caption{The disc mass (top panel) and the normalized magnetic flux (middle panel) versus the time for $M_{\rm d0}=0.1M_{\odot}$, and $q=1/2$. The solid and the dashed curves correspond to $\psi=0.5$ and $\psi=0.001$, respectively. The bottom panel shows the normalized magnetic flux versus the disc mass for $\psi=0.5$. In this figure, we consider different values of $\beta_0$, as labeled.}\label{fig:f2}
\end{figure}

In the absence of disc mass removal by wind, the total disc mass reduces only owing to the accretion onto the central star. When the contribution of the winds in mass loss is included, the disc mass reduces at a faster rate. Since a PPD lifetime $t_{\rm disc}$ can be determined in term of its total mass when it drops below a certain fraction of the initial mass, it is worthwhile to investigate the total disc mass evolution subject to the different levels of the magnetic field strength. In our work, we define $t_{\rm disc}$ as a time period that the total disc mass decreases to 1 percent of its initial mass. In the top panel of Fig. \ref{fig:f2}, evolution of the total disc mass, $M_{\rm d}$, is shown corresponding to the presented solutions in Fig. \ref{fig:f1}. Each curve is labeled with the adopted parameter $\beta_0$ and the solid and dashed curves correspond  to $\psi = 0.5$ and $0.001$, respectively. At the early phases, reduction of the disc mass with time is not significant, however, the total disc mass starts to  decline rapidly beyond a certain time which depends on the magnetic field strength. We found that disc mass reduction starts sooner when the wind is stronger. A disc with the initial mass $0.1$ $M_{\odot}$ and $\beta_0 =10^3$ will be completely depleted after about $2\times 10^4$ yr when its total mass reduces to less than one percent of its initial mass. As the wind becomes weaker, however, reduction of the disc mass occurs with a slower rate. For $\beta_0 =10^4$, the disc mass reaches about $0.01$ $M_\odot$ in $10^5$ yr, whereas for $\beta_0 =10^5$ and during the same time period, the disc mass reduces to about $0.06$ $M_{\odot}$. Using our numerical solutions, in the appendix, we present  a fitted relation between disc mass, time and the magnetic field strength for $\psi=0.01$. On the other hand, one can find that reduction of the disc mass depends on the wind parameter $\psi$. When $t=1.0$ Myr, the disc mass reaches to 0.01 $M_{\odot}$ for $\psi=0.5$ and $\beta_0=10^5$, whereas for $\psi=0.001$ and the same value of plasma parameter the disc has a mass nearly equal to 0.014 $M_{\odot}$.

\begin{table}
\begin{center}
\caption{The lifetime of a disc, $t_{\rm disc}$, for different values of $\beta_0$ and $\psi$.}\label{tab:1}
\begin{tabular}{||c|c|c||}\hline
~~ & $t_{\rm disc} (\psi=0.001) [\rm Myr]$ & $t_{\rm disc} (\psi=0.5) [\rm Myr]$ \\ \hline
$\beta_0=10^5$ & 15.3 & 10.3 \\\hline
$\beta_0=10^4$ & 0.7 & 0.5 \\\hline
$\beta_0=10^3$ & 0.016 & 0.014 \\ \hline
\end{tabular}
\end{center}
\begin{tablenotes}
   \small
   \item \textbf{Notes}: Column 1: values of $\beta_0$. Columns 2: disc lifetime for $\psi=0.001$. Column 3: disc lifetime for $\psi=0.5$.\\
\end{tablenotes}
\end{table}

The radial profile of the net vertical magnetic field strength and its evolution are not well constrained. We, therefore, considered a simple approach by assuming that magnetic flux $\Phi_{\rm B}$ is distributed so that the midplane plasma parameter $\beta_0$ remains spatially and temporally constant. We nevertheless do not impose any further constraint on the evolution of the magnetic flux, as adopted in \cite{Armitage2013} and \cite{Bai2016}. The evolution of magnetic flux $\Phi_{\rm B}$ is explored using   our disc solutions:  
\begin{align}
\nonumber \Phi_{\rm B}=2\pi \int_{r_{\rm in}}^{r_{\rm out}}B_z r dr~~~~~~~~~~~~~~~~~~~~~~\\
=(\frac{32\pi^3}{\beta_0})^{1/2}\int_{r_{\rm in}}^{r_{\rm out}} {P_{\rm mid}}^{1/2} r dr.    
\end{align}
The above equation can be written as
\begin{equation}
\Phi_{\rm B}= \Phi_{\rm B0} {m_{\star}}^{1/4}{\beta_0}^{-1/2}\int_{x_{\rm in}}^{x_{\rm out}}y^{1/2}x^{(1-q)/4}dx,
\end{equation}
where $\Phi_{\rm B0}=4\pi {r_0}^2\big(\sqrt{2\pi} \Sigma_0 \Omega_0 c_{\rm s0}\big)^{1/2}= 8.6\times10^{23} $ Wb.

In the middle panel of Fig. \ref{fig:f2}, the normalized magnetic flux evolution is shown corresponding to the solutions presented in Fig. \ref{fig:f1} and for different midplane $\beta_0$, as labeled. The wind mass loss parameter is $\psi =0.5$ (solid) and 0.001 (dashed).  One can see that the magnetic flux does not change significantly at the early phase of the disc evolution. The disc, however, is found to evolve with  constant magnetic flux  during a longer time period when the magnetic strength is weak. Rapid magnetic flux evolution is clearly found over fairly short time periods for the strong wind case. At a certain point in time, nevertheless, the decline of the magnetic flux begins and this trend is similar to the total disc mass evolution. We, therefore, display the profile of the magnetic flux versus total disc mass in the bottom panel of Fig. \ref{fig:f2} for $\psi=0.5$ and $q=1/2$. Each curve is labeled with the midplane parameter $\beta_0$. We interestingly find that our solutions correspond to a linear correlation between $\log (\Phi_{\rm B}/\Phi_{\rm B0})$ and $\log M_{\rm d}$. The slope of this linear correlation depends on the magnetic field strength. In the strong wind case, for example, the displayed curve is steeper. We find that the linear correlation is roughly independent of the adopted wind parameter. Using our solutions, in the appendix, we present an approximate relation for the magnetic flux as a function of the disc total mass for the wind parameter $\psi=0.01$. This relation indicates that $\Phi_{\rm B} \propto M_{\rm d}^{\epsilon}$ where the exponent is $\epsilon=0.39+0.09 \exp (-\beta_0 /4012)$. Therefore, the exponent $\epsilon$ varies between 0.46 and 0.39 for the allowed range of the magnetic strength, i.e. $10^3 \leq \beta_0 \leq 10^5$. Note that \cite{Bai2016} explored the  disc evolution subject with a linear relation between the magnetic flux and the total disc mass which implies a stronger correlation between these quantities in comparison to our obtained relation. Our disc solutions, however, do not correspond to an extreme situation where the magnetic flux remains constant during disc evolution.

\begin{figure}
\includegraphics[scale=1.0]{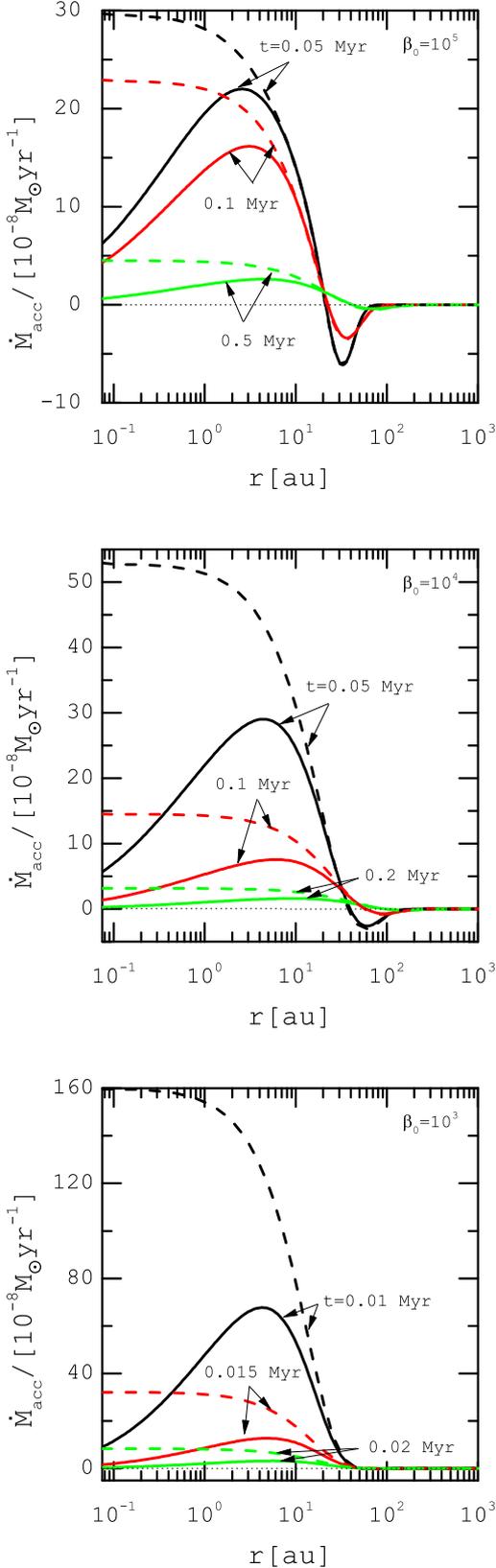} 
\caption{Profiles of the normalized accretion rate, $\dot{M}_{\rm acc}/[10^{-8}M_{\odot}{yr}^{-1}]$, versus the radial distance are shown at different times, as labeled. The initial disc mass and the exponent of temperature are assumed to be $0.1 M_{\odot}$ and $1/2$, respectively. The top panels are associated to $\beta_0=10^5$, while the middle and bottom panels correspond to $\beta_0=10^4$ and $\beta_0=10^3$, respectively. The solid and dashed curves correspond to a case with $\psi=0.5$ and $\psi=0.001$, respectively.}\label{fig:f3}
\end{figure}
\begin{figure*}
\includegraphics[scale=1.0]{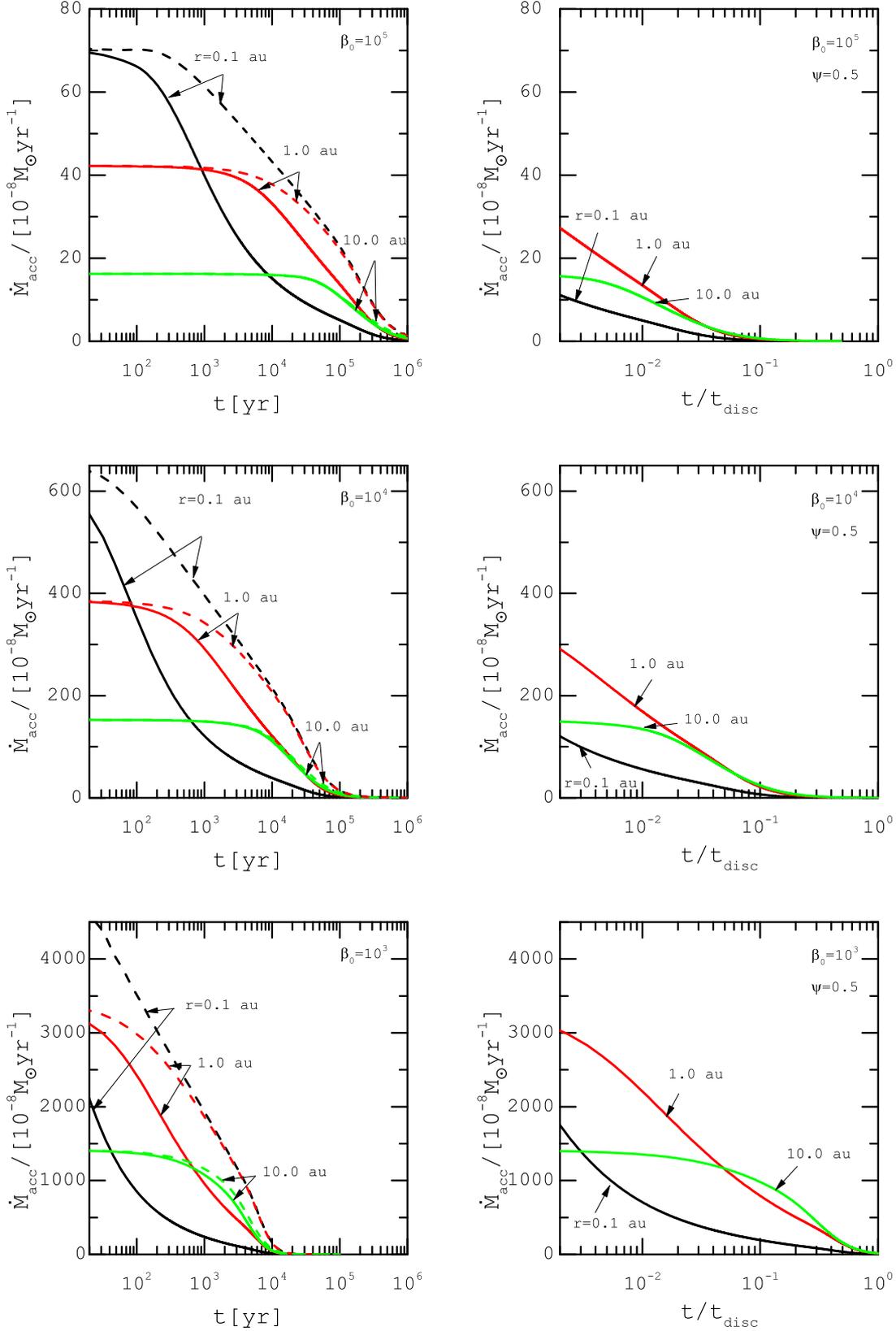}
\caption{The left panel shows profiles of the normalized accretion rate versus time for different radial distances, as labeled. The profiles of the normalized accretion rate versus the normalized time $t/t_{disc}$ for $\psi=0.5$ are also displayed in the right panel. Here, $t_{disc}$ is the disc lifetime. The initial disc mass and the exponent of temperature are assumed to be $0.1 M_{\odot}$ and $1/2$, respectively. The top plots are associated with $\beta_0=10^5$, while the middle and bottom plots correspond to $\beta_0=10^4$ and $\beta_0=10^3$, respectively.}\label{fig:f4}
\end{figure*}

\begin{figure}
\includegraphics[scale=0.5]{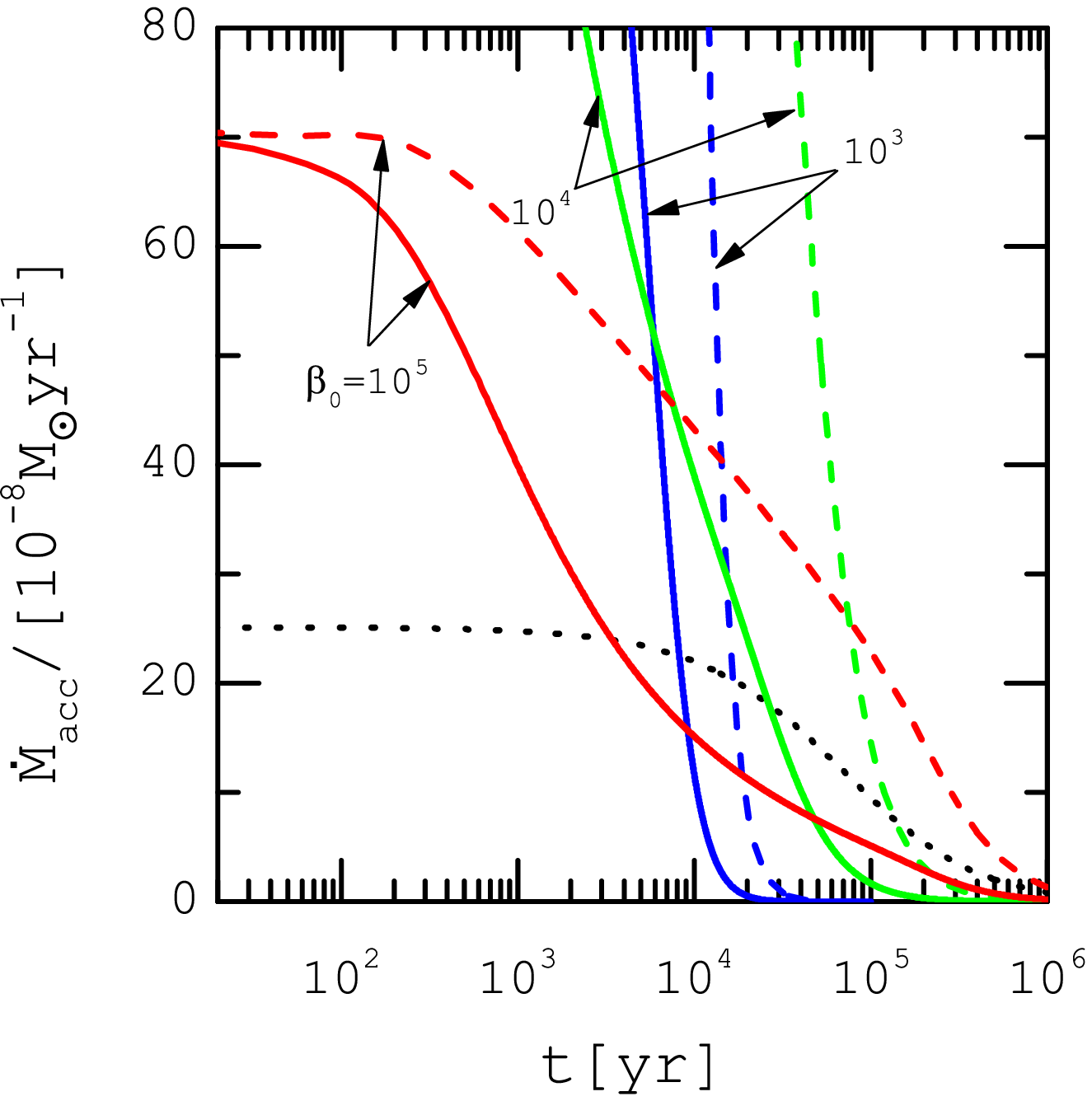}
\caption{The profile of the normalized accretion rate at $r=0.1$ au versus time for different values of $\beta_0$, as labeled. The solid and dashed curves are for $\psi=0.5$ and 0.001, respectively. The dotted curve is shown the relation of equation (\ref{eq:observ}).}\label{fig:f0}
\end{figure}

\subsection{Accretion and Wind Mass Loss}
Associated to the presented solutions in Fig. \ref{fig:f1} we can now calculate the mass accretion rate, i.e., $\dot{M}_{\rm acc}$. Note that the positive value of the accretion rate corresponds to the gas radial motion towards the central star. In Fig. \ref{fig:f3}, we show the normalized accretion rate, $\dot{M}_{\rm acc}/(10^{-8} M_{\odot} {\rm yr}^{-1})$, for different snapshots versus the radial distance and for different values of the parameter $\beta_0$. As in the previous  figures, the solid curve corresponds to $\psi=0.5$ whereas the dashed curve presents a case with $\psi=0.001$. When the magnetic wind is weak, i.e., $\beta_0 =10^5$, the accretion occurs in the inner regions whereas the gas undergoes outward motion in the disc outer parts owing to the viscous effect. 

We also find that wind mass loss significantly reduces the accretion rate in the inner part of a disc whereas the accretion rate  does not change in the outer parts. We note that in the early phase of the disc evolution, the surface density and, as a result, wind mass loss rate, is even greater, but with the time that the surface density decreases, it is expected that wind mass loss rate will be reduced too. Nevertheless, the accretion rate declines with time irrespective of the mass removal rate and the wind strength.

\begin{figure*}
\includegraphics[scale=1.0]{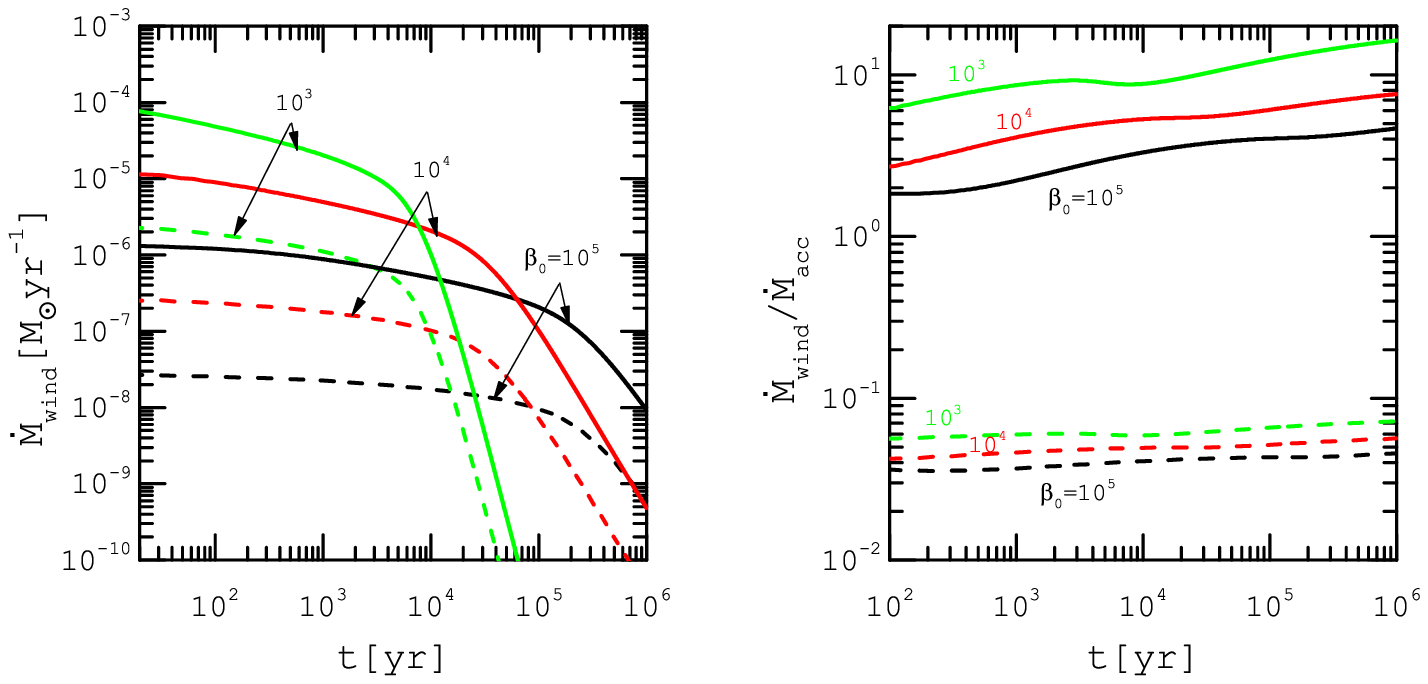}
\caption{The left panel shows the evolution of the cumulative mass loss rate by the wind, $\dot{M}_{\rm wind}$. The model parameters are $M_{\rm d0}=0.1 M_{\odot}$ and $q=1/2$ and the wind parameter is $\psi =0.5$  (solid) and 0.01 (dashed). Each curve is labeled with the corresponding parameter $\beta_0$, as labeled. In the right panel, evolution of the ratio $\dot{M}_{\rm wind}/\dot{M}_{\rm acc}$ is shown where the accretion rate $\dot{M}_{\rm acc}$ is evaluated at the radial distance 0.1 au.   }\label{fig:f5}
\end{figure*}

We now display the evolution of the normalized accretion rate at given radial distances for different magnetic field strength. In the left panels of  Fig. \ref{fig:f4}, evolution of the normalized accretion rate is shown at different radial distances, as labeled, and for $\beta_0 =10^5$ (top), $10^4$ (middle) and $10^3$ (bottom). The solid and dashed curves correspond to the cases with the wind mass loss parameter $\psi =0.5$ and 0.001, respectively. We find that the gas moves towards the star at the radial distance 0.1 au, 1 au and 10 au, irrespective of the magnetic field strength. The accretion rate, however, is significantly enhanced with increasing the parameter $\beta_0$ which means that magnetic winds contribute to the resulting inward gas flow. We also find that the accretion rate enhancement is significant at the early phase of disc evolution. In the case of weak wind, i.e., $\beta_0 =10^5$, the accretion rate at the radial distance 10 au evolves uniformly over a longer time period in comparison to the cases with moderate wind ($\beta_0 =10^4$) or strong wind ($\beta_0 =10^5$). But the rapid decline of the accretion rate is found beyond a certain time. This typical behavior is found at other radial distances no matter what value is adopted for the wind mass loss parameter. Obviously, the accretion rate decreases further as the wind mass loss parameter $\psi$ increases. However, the difference between the accretion rates in the cases with $\psi=0.001$ and $\psi =0.5$ becomes more significant in the disc inner part. Note that the inner disc surface density evolution exhibited  similar behaviour as we explored earlier. Thus, wind mass loss plays a crucial role  in the inner disc evolution.

In the right panels of Fig. \ref{fig:f4}, the normalized accretion rate versus the normalized time is shown for $\psi=0.5$. As we mentioned before, the disc lifetime  $t_{\rm disc}$ is defined as when the total disc mass decreases to 1 percent of its initial mass. The lifetime of a disc with $\beta_0=10^3$ and $10^4$ is found as 0.014 and 0.5 Myr, respectively. We also find that the lifetime of a PPD with $\beta_0=10^5$ is about 10.3 Myr which exceeds the expected normal lifetime of the PPDs. The accretion rate significantly increases with the magnetic field strength, however, this enhancement in the inner region is more pronounced. In the weak wind case, for instance, we find that the inner accretion rate at the radial distance 0.1 au is about $ 10^{-7} $ M$_{\odot}$yr$^{-1}$ whereas this rate increases with a factor 10 for $\beta_0 =10^4$, and in the strong wind case, the inner accretion rate significantly increases to about $1.7\times 10^{-5}$ M$_{\odot}$yr$^{-1}$. But the accretion rate at a larger radial distance, say 1 au or 10 au, increases approximately proportional to the midplane parameter $\beta_0$. In the weak and intermediate wind cases, however, we find that significant decline of the accretion rate occurs during 10 percent of the disc lifetime, whereas in the strong wind case, reduction of the accretion rate continues during the entire lifetime of the disc. In other words, the contribution of the magnetic winds in the accretion rate is expected to persist even in the evolved PPDs.

We can now compare our evolutionary profile of the accretion rate with an observationally motivated relation for this quantity \citep{Hartmann1998,Bitsch2015,Izidoro2019}. The accretion rate as a function of the disc age $t$ is written as  
\begin{equation}\label{eq:observ}
    \log\big(\frac{\dot{M}_{\rm acc}}{M_{\odot} {\rm yr}^{-1}}\big)=-8-1.4\log\big(\frac{t+10^5 {\rm yr}}{10^6 {\rm yr}}\big),
\end{equation}
This relation is illustrated in Fig. \ref{fig:f0} by the dotted curve. Therefore, it is reasonable to expect the total wind mass loss to gradually decrease as the disc evolves and becomes older. We display evolution of the normalized accretion rate at the radial distance $0.1$ au  for different values of $\beta_0$, as labeled. The solid and dashed curves correspond to the cases with $\psi=0.5$ and 0.001. The general trend of the above mentioned relation for the accretion rate is qualitatively similar to our obtained behaviors, however, our model predicts a larger accretion rate at the early phase of disc evolution. This finding implies that a single relation for the accretion rate evolution does not adequately describe its behavior because the magnetic wind strength and the wind mass loss rate profoundly affect the evolution of the accretion rate.  

The wind mass loss rate $\dot{M}_{\rm wind}$ is calculated using equation (\ref{eq:Mwind}) and its evolution is depicted in the left panel of Fig. \ref{fig:f5} for $\psi=0.5$ (solid) and 0.01 (dashed) and different values of $\beta_0$, as labeled. Other model parameters are same as in Fig. \ref{fig:f1}. Note that equation (\ref{eq:Mwind}) is used to calculate the total wind mass loss rate. In the early phase of the disc evolution, the total wind mass loss rate is more or less uniform, but beyond a certain time, the wind mass loss rate started to decline. In the weak wind case, i.e., $\beta_0 =10^5$, decline of $\dot{M}_{\rm wind}$ occurs at about $2\times 10^5$ yr. As the wind becomes stronger, the decline of the wind mass loss rate happens sooner, but with a faster rate. In the strong wind case with $\beta_0 = 10^3$, for instance,  the rapid decline of the total wind mass loss rate occurs at a time about $10^4$ yr. This trend, nevertheless, is more or less independent of the adopted wind mass loss parameter $\psi$. 

We thereby recognize two phases of mass removal by the magnetically winds. While at the early times of the disc evolution, the wind mass loss smoothly decreases with time, the disc thereafter undergoes a rapid reduction in the wind mass loss rate. We think this trend is the outcome of inside-out reduction of the disc surface density due to the presence of winds. In our model, magnetic winds contribute to the angular momentum removal and mass loss and both these rates are directly proportional to the disc surface density. At the early phase of the disc evolution, we see that only the very inner part with the highest density is affected by the wind whereas reduction of the surface density in rest of the disc is very negligible. As time proceeds, a larger portion of the disc exhibits surface density reduction. Once a large enough part of the disc is affected by the winds, rapid declines of the disc quantities such as its total mass and magnetic flux are observed. As the wind gets stronger, the onset time of the rapid decline occurs sooner which means that the duration of the initial slow decline is shorter as a  smaller value of parameter $\beta_0$ is adopted.

In the right panel of Fig. \ref{fig:f5}, evolution of the ratio $\dot{M}_{\rm wind}/\dot{M}_{\rm acc}$  is shown where the accretion rate $\dot{M}_{\rm acc}$ is evaluated at the inner radius 0.1 au. As before, the solid  curves stand for a case with $\psi=0.5$ and the dashed curves correspond to a case with $\psi=0.01$. The ratio $\dot{M}_{\rm wind}/\dot{M}_{\rm acc}$ increases with time and this enhancement is amplified with the magnetic field strength. This trend implies that wind mass removal gradually becomes larger than the inner accretion rate despite its rapid decline during the final phase of the disc evolution. In the weak wind case with $\psi=0.01$, for instance, wind mass loss rate reaches to about 4.5 percent of the inner accretion rate during $10^6$ yr, whereas in the strong wind case, this factor is about 7 percent during the same time period. With a high wind mass loss parameter, i.e., $\psi =0.5$, the wind mass loss rate is generally larger than the inner accretion rate which is unlikely to be supported by the observed PPDs. We thereby consider $\psi=0.01$ as a standard value for the wind parameter in the next figures.

\subsection{Disc Size Evolution}

\begin{figure}
\includegraphics[scale=1.0]{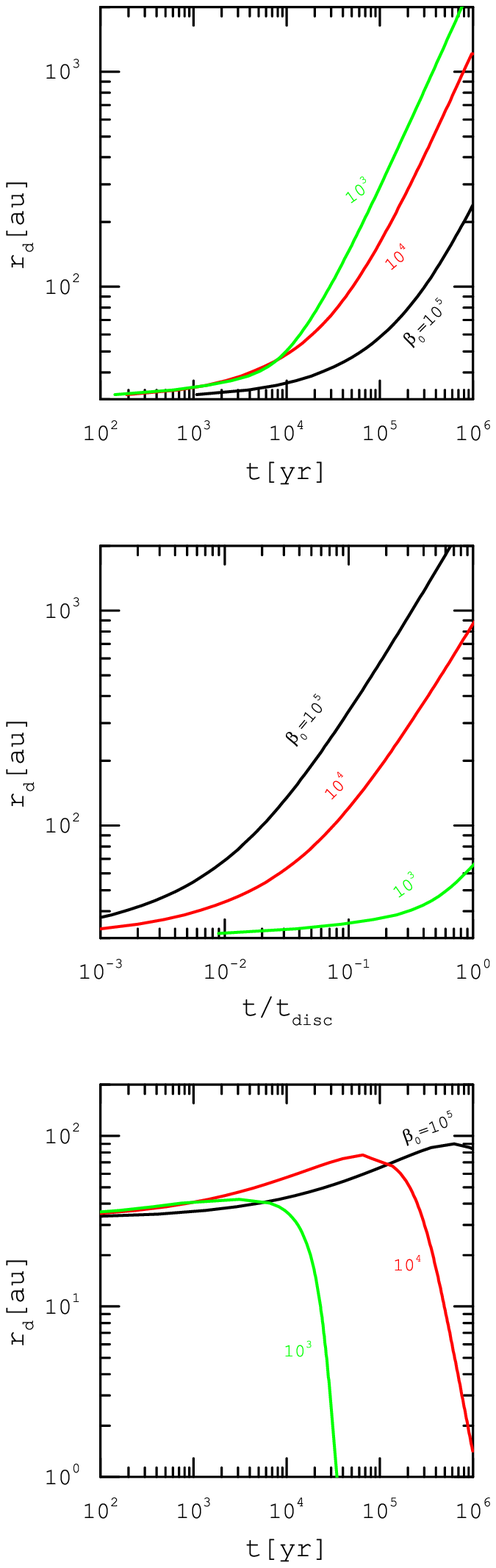}
\caption{The disc radius for $\psi=0.01$ and $q=1/2$. In the top and middle panels, the disc radius is defined using the enclosed mass. In the bottom panel, however, we have benefited from threshold surface density for defining the disc radius.}\label{fig:f6}
\end{figure}

In the panels of Fig. \ref{fig:f6}, the disc radius as a function of time is shown for different values of $\beta_0$ using different definitions for a disc size. In the top panel of Fig. \ref{fig:f6}, we define disc radius as a radial distance where the enclosed mass, $M_{\rm enc}$, is a given fraction, $\chi$, of the total disc mass: $M_{\rm enc}(r_{\rm d}t )=\chi M_{d}(t)$. Note that our results are qualitatively independent of the adopted fraction $\chi$. For simplicity, we assume that the fraction is $\chi=0.99$. This disc size definition has already been implemented by some authors \citep[e.g.,][]{Anderson2013,shadmehri2018}, but as we discuss now, it does not properly address the expected disc size evolution in the presence of the magnetic winds.   

The top panel of Fig. \ref{fig:f6} shows that disc size increases with time and in the presence of  wind the disc spreading occurs with a faster rate. This trend appears to contradict our prior arguments that magnetic winds are able to suppress disc spreading which occurs due to the viscous effects. But it is a direct consequence of the adopted disc size definition. As the wind becomes stronger, the disc mass reduces with a faster rate and we have to include a larger portion of a disc to account for the enclosed mass as a given fraction of the total mass. In the middle panel, we plot disc size as a function of the normalized time, i.e., $t/t_{\rm disc}$. Now, the trend is reversed and as the wind becomes stronger, the disc spreading occurs with a slower rate. Note that these behaviors are found qualitatively independent of the adopted wind mass loss parameter.  

We now define disc size as a radius where the surface density becomes a given threshold value.   We adopt a fixed  threshold value 1.0 g ${\rm cm}^{-2}$ for determining radius of the studied discs. In  the bottom panel of Fig. \ref{fig:f6}, we plot disc size evolution using this new definition. The disc undergoes a radial expansion due to the viscous effects at the early phase of the evolution. While in the weak wind case, disc spreading continues more or less over the entire disc lifetime, this size expansion eventually is halted and size shrinkage occurs in the intermediate and strong wind cases. We thereby expect that evolved PPDs to have a smaller size in comparison to a less evolved system when magnetic winds play a significant role.

\begin{figure}
\includegraphics[scale=1.0]{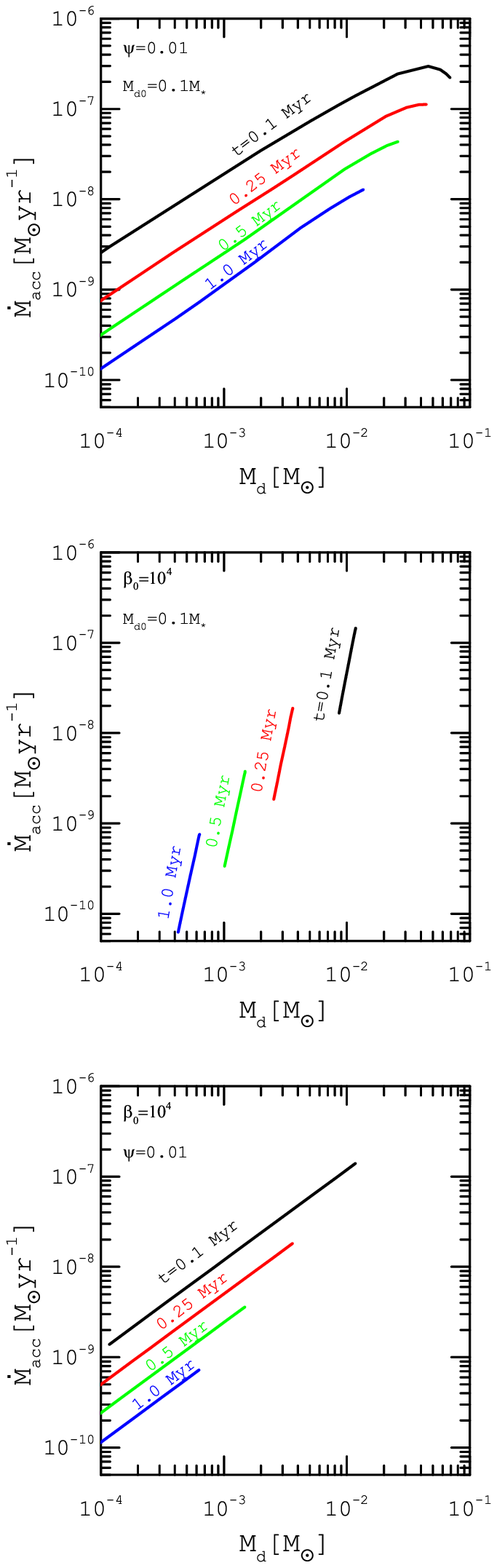}
\caption{Isochrone tracks of an ensemble   of  PPDs at different ages, as labeled. The accretion rate is evaluated at the radius 0.1 au and all cases correspond to a solar mass central star. The top panel shows the role of $\beta$ on the isochrone tracks within its permitted range, i.e., $10^{3} \leq \beta \leq 10^5$.  The influence of $\psi$ on these quantities is displayed in the middle panel. It is assumed that $\psi$ varies from 0.001 to 0.5. The bottom panel shows the effect of the initial disc mass on the accretion rate and the disc mass. The range of $M_{\rm d0}$ is 0.001-0.1$M_{\star}$. }\label{fig:f7}
\end{figure}

\subsection{Isochrone Tracks for PPDs with Winds}
We can reconcile our disc evolution model with the observed PPDs using isochrone concept which has been implemented by LSMT17 in the context of PPD modeling. The isochrone tracks in the accretion rate and total disc mass plane for an ensemble of PPDs specify locus of points corresponding to the discs with the same age and the same initial disc mass. LSMT17 obtained the isochrone tracks using similarity solutions with viscosity as a power-law function of the radius. They found that
\begin{equation}\label{eq:iso}
    \dot{M}_{\rm acc}=\frac{M_{\rm d}}{2(2-\gamma ) t} \left [ 1- \left(\frac{M_{\rm d}}{M_{\rm  d0}} \right )^{2(2-\gamma )}\right ],
\end{equation}
where $\gamma$ is the viscosity exponent and $t$ denotes the disc age. In the absence of the winds, we verified that the above analytical isochrone tracks are obtained by solving our main equation (\ref{eq:main1}) numerically. But we now aim to explore how the isochrone tracks are modified when magnetically driven winds contribute to the angular momentum removal and disc mass loss.

Figure \ref{fig:f7} displays isochrone tracks for PPDs with one solar mass central star and different disc ages, as labeled. The accretion rate $\dot{M}_{\rm acc}$ is evaluated at the inner radius 0.1 au. All tracks in the top panel of Fig. \ref{fig:f7} correspond to the discs with the same initial disc mass, i.e., $M_{\rm d0}=0.1$ $M_{\star}$, and the wind parameter is $\psi=0.01$. The tracks are truncated because the allowed range of the magnetic strength is between $\beta=10^3$ and $10^5$. At the early phase of the evolution, the tracks are more or less resemble the viscous disc tracks represented by equation (\ref{eq:iso}) because there is not enough time for the wind to affect disc structure significantly. At the time $t=0.1$ Myr, for instance, the accretion rate increases with the disc mass, but it starts to decline beyond a certain disc mass. Reduction of the accretion rate with the disc mass, however, is not found as time proceeds owing to the winds. But the accretion rate increases as a power-law function of the disc mass with an exponent of about 1.0 irrespective of the disc age. 

In the middle panel of Fig. \ref{fig:f7},  role of the wind mass loss parameter $\psi$ on the isochrone tracks is illustrated for $\beta_0=10^4$ and $M_{\rm d0}=0.1$ $M_{\star}$. The range of the wind parameter is assumed to be $0.001-0.5$. We find a power-law relation between the accretion rate and the disc mass where its exponent is dependent on the disc age. This exponent  reduces as the disc age increases. For example, the exponent at the times $t=0.1$ Myr and $t=1.0$ Myr is about 6.9 and 6.2, respectively. We also study the influence of initial disc mass in the bottom panel of Fig. \ref{fig:f7}. In this case, we consider $\beta_0=10^4$ and $\psi=0.01$. It is assumed that $M_{\rm d0}$ varies from 0.001 $M_{\star}$ to 0.1 $M_{\star}$. We find a power-law relation between that $\dot{M}_{\rm acc}$ and $ M_{\rm d}$ with an exponent equal to 1 no matter what is disc age.

\begin{figure*}
\includegraphics[scale=1.0]{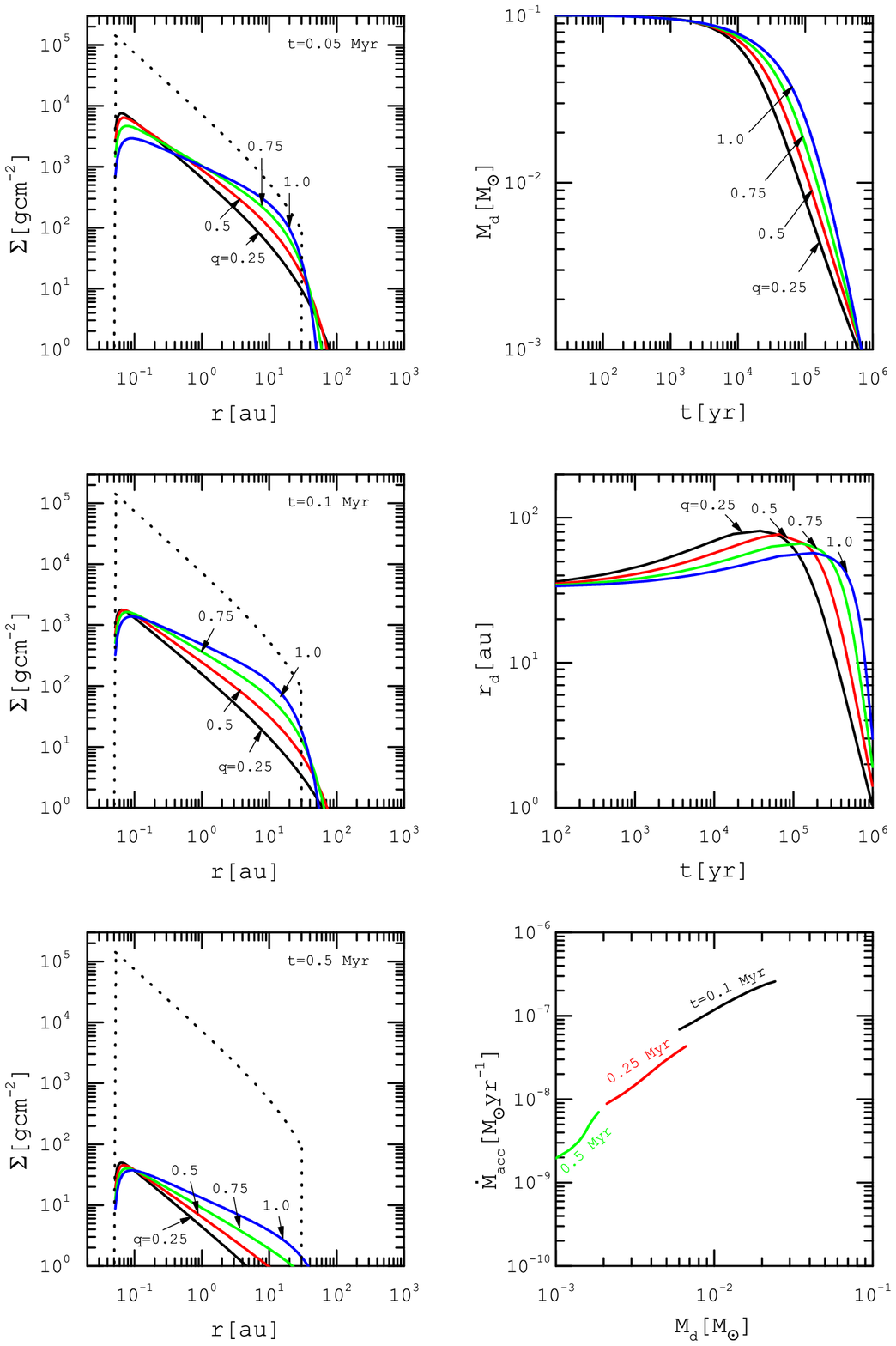}
\caption{Effect of the temperature exponent $q$ in the surface density  (left panels), the disc mass  (right, top), the disc size (right, middle) and the isochrone tracks (right, bottom) are explored for the discs with one solar mass central star and  $\beta_0=10^4$ and $\psi=0.01$. In the left panels, each curve is labeled with the temperature exponent. The isochrone tracks (right, bottom) are obtained using a more extended range of the temperature exponent  between 0.1 and 1 and for the discs with different ages, as labeled. The accretion rates are evaluated at radius of 0.1 au.}\label{fig:f8}
\end{figure*}

\subsection{Role of Temperature Exponent}\label{sec:q}
In all the figures we have presented so far, the temperature exponent was $q=1/2$, but we now investigate the role of this parameter in the disc evolution and the associated isochrone tracks. The left panels of Fig. \ref{fig:f8} display surface density evolution for the discs with one solar mass central star and different temperature exponent and different times, as labeled. The initial surface density distribution is shown by a dashed curve. The initial disc mass, as before, is $M_{\rm d0}=0.1 M_{\star}$ and the magnetic strength and the wind parameter are $\beta_0 = 10^4$ and $\psi =0.01$. Note that our adopted power-law  temperature profile becomes steeper with increasing its exponent $q$. The disc surface density decreases with time, however, its radial distribution becomes steeper as the temperature exponent $q$ decreases. While surface density evolution at the early times slightly depends on the adopted exponent $q$, as time proceeds, the evolution of the surface density at the very inner part becomes more or less independent of the temperature exponent. A disc with a lower temperature exponent, thereby, is dispersed over a shorter time period. 

In the top right  panel of Fig. \ref{fig:f8}, the evolution of the total disc mass is shown for different temperature exponent, as labeled. The total disc mass does not undergo significant reduction at the early times irrespective of the adopted parameter $q$. The rapid decline of the total disc mass is started after about $10^4$ yr, however, its rate is faster for the cases with a lower $q$. Although the disc lifetime can be slightly enhanced with increasing the exponent $q$, this dependence is not very significant. For adopted values of $q$ in this figure, we find that disc lifetime is between 0.6 Myr and 0.7 Myr.  In the middle right panel of Fig. \ref{fig:f8}, we display disc size evolution using threshold density criterion and for different temperature exponent, as labeled. The disc radius initially increases with time, however, this radial expansion is halted after a certain time period and the disc starts to shrink. The radial expansion occurs sooner and with a faster rate as the temperature exponent decreases. The associated isochrone tracks are shown in the bottom right panel of Fig. \ref{fig:f8} for the ages 0.1, 0.25, and 0.5 Myr. Each curve is labeled with the corresponding disc age and the temperature exponent is permitted to vary between 0.1 and 1.0. The tracks can be approximated using a linear relation between $\log \dot{M}_{\rm acc}$ and $\log M_{\rm d}$ where its slope varies from 0.98 to 2.1 for the ages 0.1 Myr and 0.5 Myr, respectively.

\begin{figure*}
\includegraphics[scale=1.0]{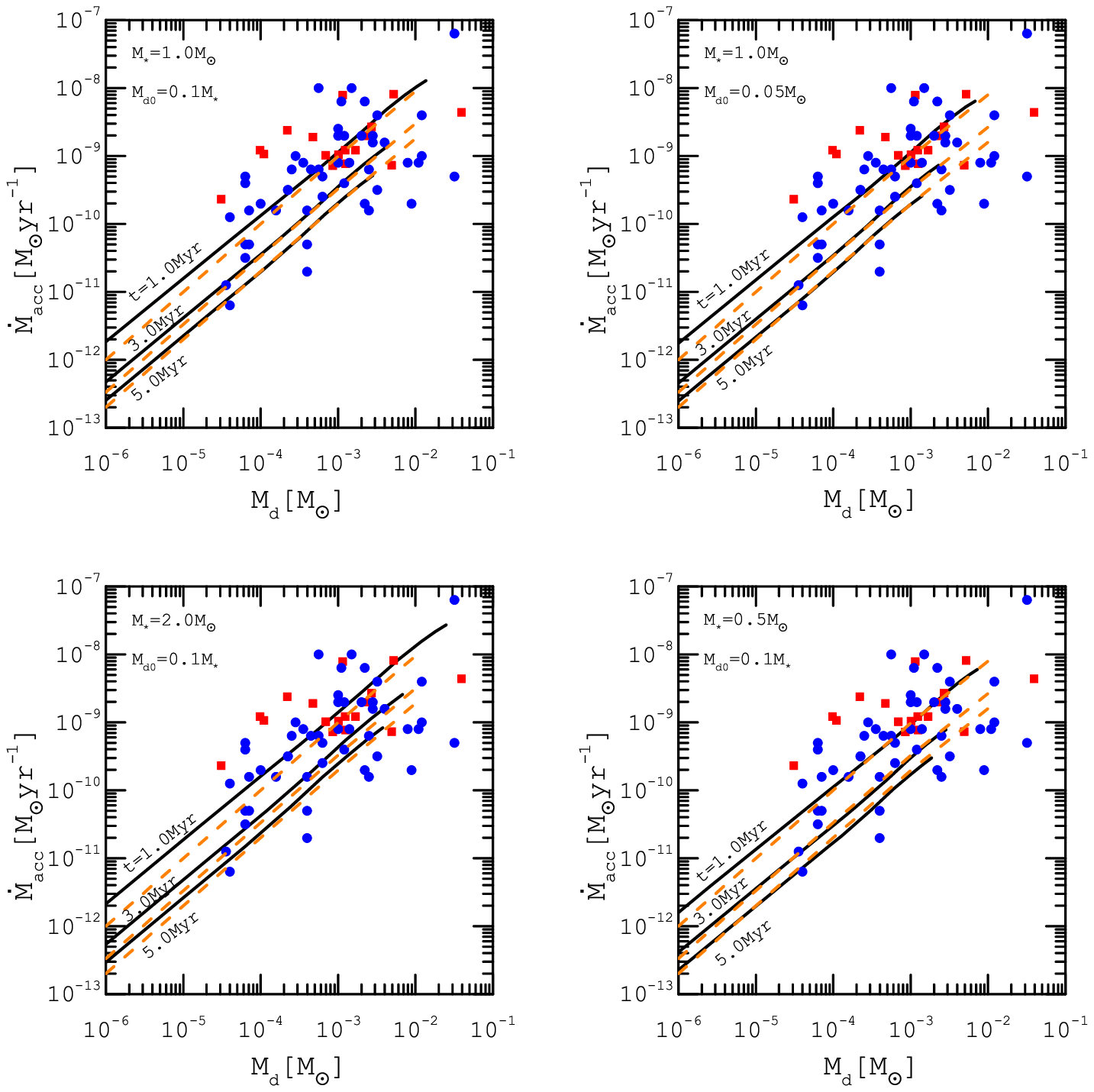}
\caption{The influence of $\beta_0$ on profiles of $\dot{M}_{\rm acc}-M_{\rm d}$ of a PPD at ages=1.0, 3.0, and 5.0 Myr, as labeled, along with the observational data from the $\sigma-$Orion \citep{Ansdell2017, Rigliaco2011} and the Lupus \citep{Manara2016} survey. The observational data of $\sigma-$Orion and Lupus are shown with red squares and blue circles, respectively. The orange dashed lines correspond to the results of a viscous model for $\gamma=3/2$ and ages=1.0, 3.0, and 5.0 Myr (LSMT17). We assume that the wind parameter $\psi$ is 0.01. The magnetic strength $\beta_0$ can also range from $10^3$ to $10^5$. }\label{fig:f9}
\end{figure*}

\section{astrophysical implications}

We can now reconcile our theoretical results with the observed PPDs to understand if these trends are supported by the observations. Isochrone tracks, in particular, are a very useful tool for such a study. A similar comparison, as we mentioned earlier, has been done by LSMT17 using a purely viscous disc model in the absence of the magnetic winds. Our model thereby generalizes this prior by incorporating both viscous effects and the angular momentum and mass removals by the magnetic winds. In doing so, we use the results of  \cite{Ansdell2017} and \cite{Rigliaco2011} for $\sigma-$Orion, and the results of \cite{Manara2016} for Lupus. In  Fig. \ref{fig:f9}, we display theoretical tracks within the allowed range of the parameter $\beta_0$ by solid curves and different ages, as labeled. The observational data of $\sigma-$Orion and of Lupus are shown by red squares and blue circles, respectively. \cite{Manara2016} and \cite{Ansdell2017} reported that the mass of the central star is within $(0.1-2.2) M_{\odot}$. Our theoretical tracks, therefore, correspond to the stellar mass from $0.5 M_{\odot}$ to $2.0 M_{\odot}$, as labeled in each panel. For easier comparison of the tracks, we also plot theoretical tracks of LSMT17 using equation (\ref{eq:iso}) by the orange dashed lines. In all the panels of Fig. \ref{fig:f9}, the wind mass loss parameter is $\psi=0.01$, because as we argued earlier, in the discs with a higher $\psi$ the wind mass loss rate becomes much larger than the inner accretion rate which is unlikely to be consistent with the observed  PPDs. The slope of our theoretical isochrone tracks is found independent of the initial disc mass, whereas a purely viscous disc model predicts a slope with a strong dependence on the initial disc mass (LSMT17). We also find that the slope of the tracks increases with the mass of the central star.     

One can see that the best fit for Lupus is obtained for $t=$1.0 Myr, well within the estimated age of this star forming region, i.e., $(1.0-3.0)$ Myr. As we have found earlier, the slope of the isochrone track slightly depends on the host star mass. This slope at the age 1.0 Myr, for instance, is found 0.93 and 0.95 for the host star mass $M_{\star}=0.5 M_{\odot}$ and $2.0 M_{\odot}$, respectively. LSMT17 found that the slope is about 1 when the disc mass is low.  We also find that the best-fit $\beta_0$ value depends on the input parameters. For  $M_{\star}=1.0 M_{\odot}$ and $M_{\rm d0}=0.05 M_{\star}$, the best-fit $\beta_0$ value is within the range $6\times10^3-6\times10^4$. However, the best-fit $\beta_0$ value can range from $5\times10^3$ to $3\times10^4$ for a PPD with $M_{\star}=2.0 M_{\odot}$ and $M_{\rm d0}=0.1 M_{\star}$.

For $\sigma-$Orion, the best fit is obtained between $(1-3)$ Myr. If $M_{\star}=1.0 M_{\odot}$ and $M_{\rm d0}=0.05 M_{\star}$, the best-fit $\beta_0$ value is within the range  $(1-4)\times10^4$. However, this range becomes $8\times10^3-2\times10^4$ if we adopt $M_{\star}=2.0 M_{\odot}$ and $M_{\rm d0}=0.1 M_{\star}$. For  $M_{\star}=0.5 M_{\odot}$, the slope of the isochrone track becomes 0.949 and 0.953 at the ages 3.0 Myr and 5.0 Myr, respectively. These values for $M_{\star}=2.0 M_{\odot}$ are modified to 0.963 and 0.966 at ages 3.0 Myr and 5.0 Myr. Note that the observational results related to this region lead to a slope of about 0.8 \citep{Ansdell2017, Rigliaco2011}. We also note that ALMA surveys of nearby star-forming regions have shown that the dust mass in the disk is correlated with the stellar mass, but with a large scatter. This scatter could indicate either different evolutionary paths of disks or different initial conditions within a single cluster.

\section{discussion}
We studied disc evolution with magnetically driven winds using relations for the rates of the angular momentum removal and the wind mass loss rate motivated by the recent non-ideal MHD disc simulations \citep{Simon2013b,Hasegawa2017}. We then explored the evolution of the disc quantities subject to a wide range of the model parameters. As we mentioned before, a similar model has also been implemented by \cite{Armitage2013}, however, they neither investigated the evolution of the disc quantities such as the total disc mass, the accretion rate and the associated isochrone tracks nor incorporated the mass loss by the wind. Other previous models of PPDs with the magnetic wind also did not explore isochrone tracks \citep[e.g.,][]{Bai2016,Suzuki2016,Ogihara2018}. Considering recent advances which highlight the importance of the wind driven accretion, we  think that our model provides further physical insights about wind driven accretion paradigm and its predictions that can be confronted with the observed PPDs. 

Providing theoretical robust diagnostics to discriminate between various accretion processes in PPDs is still a challenging issue. To our knowledge, current studies generally follow four strategies to address this essential problem: (i) Determining disc size can be used to understand whether magnetic winds play a dominant role in the evolution of PPDs \citep{Najita2018}. (ii) Level of the observed accretion  in a few well-studied PPDs apparently can not be explained using MRI driven accretion and magnetic wind driven accretion is  plausible \citep{Wang2017}. (iii) The inferred value of the well known $\alpha$ parameter spans a wide range from 0.0001 to 0.04 and probably MHD winds are responsible for decoupling of the central accretion rate from the global accretion rate \citep{Rafikov2017}. (iv) Location of the wind emitting region is less than 1 au (i.e., inside the gravitational radius for thermal escape) in 65 Tauri systems which point to consider  MHD winds \citep{Fang2018,Banzatti2019}. We, however, note that all the above mentioned studies deserve further studies to obtain conclusive results. Our study nevertheless proposes that isochrone concept is a  diagnostic that can be added to the above list. 

Aside from the above issue, we also found  evolutionary features that can occur commonly in the PPDs with magnetically driven winds. The inner region of a PPD is affected by the wind at the early phase of the disc evolution, but as time proceeds, we can see wind effects in the other disc regions. The disc surface density reduction due to wind launching proceeds with an "inside-out" fashion irrespective of the adopted model parameters. We quantified this trend in terms of the magnetic wind strength and found that in the strong wind case with $\beta_0 =10^3$ reduction in the disc surface density and the corresponding total disc mass occurs during a time period much shorter than the   PPDs lifetimes motivated by the observations. Our model, therefore, reinforces an earlier theoretical expectation that MHD winds are launched mostly from the inner parts of PPDs where the photoevaporative winds are not efficient. Whether the observed winds or outflows in some of the observed PPDs are driven by magnetic fields or radiation force can be addressed by spatially resolving the wind emitting region. 

We also found that discs with the magnetic winds generally undergo two evolutionary phases no matter what the model parameters are chosen. During the first phase of evolution, global disc quantities such as disc total mass and the magnetic flux smoothly decrease with time, though other disc quantities including surface density exhibit significant reduction within this time interval. The second phase of evolution is started with a rapid decline in the global disc quantities. Duration of each phase, however, depends upon the model parameters. In the weak (strong) wind case, for instance, the total disc mass does not reduce significantly during the first $10^4$ yr ($10^2$ yr) of its evolution and rapid mass decline occurs thereafter. The magnetic flux, on the other hand, does not exhibit  noticeable change within the first $10^5$ yr ($10^3$ yr) of the evolution in the weak (strong) case. These trends suggest that contribution of the magnetic winds in the evolution of very young PPDs is negligible, however, the subsequent phase of the evolution is largely dominated by the magnetic winds. This second phase nevertheless may last a significant fraction of PPDs lifetimes. We, however, note that more sophisticated disc models are needed to determine whether or not this two-phase evolutionary scenario occurs in the real PPDs.

We also calculated the accretion rate at different radial distances in the presence of the magnetic winds. The theoretical profile was then compared with an observationally motivated relation introduced by \cite{Hartmann1998}. Although the general trends of both profiles are similar, we could not find quantitative agreement unless at the late stage of evolution. Our model predicts that the accretion rate at the inner region largely depends upon on the wind strength and the mass loss rate parameter. It is therefore unlikely to describe the accretion evolution for an ensemble of PPDs using a single approximate relation. 

The accretion rate at  the inner region, however, is largely dominated by the magnetic winds for $\beta_0 < 10^4$. This state corresponds to the wind driven accretion scenario, in which the angular momentum transport is suppressed and the accretion primarily occurs due to the angular momentum removal by the magnetic winds. This process effectively reduces disc lifetime. We also found that the cumulative mass loss rate by wind also increases with the magnetic wind strength. These findings imply that magnetic winds can efficiently contribute to the disc dispersal and shorten the lifetime. This mechanism is restricted to the inner disc region at the early phase of the disc evolution, but as time proceeds, the entire structure of the disc can be affected by wind launching. This trend raises an important question regarding  the role of the photoevaporative winds in the outer disc parts in the presence of the magnetic winds. Dispersal of PPDs is commonly attributed to the photoevaporative winds, however, our present study suggests that magnetic winds can also play a significant role in shortening PPDs lifetimes. It is, therefore, an interesting problem to explore evolutions of the discs in the presence of both photoevaporative and magnetic winds. The cumulative mass loss rate by wind also increases with wind strength. Its rate, however, rapidly decreases with time in the second phase of disc evolution. But the ratio of wind mass loss rate and the accretion rate slightly increases with time. We, therefore,  expect that magnetic winds may carry less amount of mass as the host PPD gets older.   
 
Disc size, as we discussed earlier, can serve as a good diagnostic to determine whether PPDs are subject to the MRI-driven mechanism or MHD wind-driven accretion. In the MRI driven scenario, the disc undergoes radial expansion, whereas in the wind dominated case, size expansion is strongly suppressed because magnetic wind contributes to the angular momentum removal rather than redistributing this quantity as MRI does. Our model also clearly exhibits this trend in  weak and strong wind cases. However, the implemented definition for the disc size plays a crucial role in determining the size and its evolution. 

Our numerical solutions enabled us to follow the associated tracks in the accretion rate and total disc mass plane for an ensemble of PPDs at different ages. A similar analysis has also been done by LSMT17, but in the absence of contributions of the magnetic winds. We then confronted our theoretical isochrone tracks with PPDs in the star forming regions Lupus and $\sigma$-Orion. The slope of our relation is almost 0.95. However, LSMT17 obtained a slope nearly equal to 1. Note that this slope is almost 1 in Lupus and it is 0.8 in $\sigma$-Orion. Our isochrone tracks  depend upon the age and the host star mass, whereas the correlation of LMS17 is independent of the stellar mass. We found that the slope of isochrone tracks increases with the host star mass. Although it is difficult to distinguish observationally whether a PPD undergoes solely viscous evolution or a combination of this process and angular momentum removal by magnetic winds, at least our model illustrates that the current knowledge about PPDs based on the isochrone concept is consistent with a model in which magnetic winds contribute to the disc accretion.

\section*{Acknowledgements}
We would like to than referee for a constructive report that helped us to improve the manuscript. 
This work has been supported financially by Research Institute for Astronomy \& Astrophysics of Maragha (RIAAM) under research project NO. 1/5440-13.

\section*{appendix}
In the top panel of Fig. \ref{fig:f2}, the total disc mass evolution was shown for $q=1/2$, $M_{\rm d0}=0.1 M_{\odot}$, and $\psi=0.001$ and 0.5. For $\psi=0.01$, the decline of the total disc mass begins after about 10000 yr and the following approximate relation is obtained using our numerical solutions:
\begin{align*}
 \nonumber \log M_{\rm d}=\Big[0.13\exp\big(-\frac{\beta_0}{93285.76}\big)+0.22\exp\big(-\frac{\beta_0}{1486.29}\big)\\
 \nonumber -0.03\Big](\log t)^3 +\Big[-2.85\exp\big(-\frac{\beta_0}{184922.05}\big)\\
 \nonumber -2.9\exp\big(-\frac{\beta_0}{1553.6}\big)+1.19\Big](\log t)^2\\
 \nonumber +\Big[1.26\times10^{-18}{\beta_0}^4-2.75\times10^{-13}{\beta_0}^3\\
\nonumber +1.95\times10^{-8}{\beta_0}^2-5.47\times10^{-4}\beta_0 +10.97\Big](\log t)\\
 \nonumber -9.34\times10^{-19}{\beta_0}^4+1.97\times10^{-13}{\beta_0}^3\\
-1.3\times10^{-8}{\beta_0}^2+3.32\times10^{-4}\beta_0-13.28,
\end{align*}
where $M_{\rm d}$ and $t$ are expressed in "$M_{\odot}$" and "yr", respectively.

As mentioned previously, $ \Phi_{\rm B}$ increases approximately as a power law function of the total disc mass $M_{\rm d}$. If $\Phi_{\rm B}$ and $M_{\rm d}$  are expressed in "Wb" and  "$M_{\odot}$", the following fitted relation for $\psi=0.01$ and $q=1/2$ is obtained
\begin{align*}
\nonumber  \log \Phi_{\rm B}=\Big[0.09\exp\big(-\frac{\beta_0}{4012.42}\big)+0.39]\log M_{\rm d}\\
+0.98\exp\big(-\frac{\beta_0}{13225.55}\big)+23.27.
\end{align*}

If we define the disc radius based on the enclosed mass, the evolution of $r_{\rm d}$ for a case with $\psi=0.01$ and $q=1/2$ may be written as 
\begin{align*}
 \nonumber \log r_{\rm d}=\Big[0.01\exp\big(-\frac{\beta_0}{11982.16}\big)+0.17\Big](\log t)^2+\Big[-7.79\\
 \nonumber \times10^{-16}{\beta_0}^3+1.62\times10^{-10}{\beta_0}^2-1.12\times10^{-5}\beta_0 \\
 -0.97\Big](\log t)-1.12\exp\big(-\frac{\beta_0}{25094.09}\big)+3.85,
\end{align*}
where $r$ and $t$ are expressed in "au" and "yr", respectively.
\bibliographystyle{mnras}
\bibliography{reference} % if your bibtex file is called example.bib

\begin{thebibliography}{}
\makeatletter
\relax
\def\mn@urlcharsother{\let\do\@makeother \do\$\do\&\do\#\do\^\do\_\do\%\do\~}
\def\mn@doi{\begingroup\mn@urlcharsother \@ifnextchar [ {\mn@doi@}
  {\mn@doi@[]}}
\def\mn@doi@[#1]#2{\def\@tempa{#1}\ifx\@tempa\@empty \href
  {http://dx.doi.org/#2} {doi:#2}\else \href {http://dx.doi.org/#2} {#1}\fi
  \endgroup}
\def\mn@eprint#1#2{\mn@eprint@#1:#2::\@nil}
\def\mn@eprint@arXiv#1{\href {http://arxiv.org/abs/#1} {{\tt arXiv:#1}}}
\def\mn@eprint@dblp#1{\href {http://dblp.uni-trier.de/rec/bibtex/#1.xml}
  {dblp:#1}}
\def\mn@eprint@#1:#2:#3:#4\@nil{\def\@tempa {#1}\def\@tempb {#2}\def\@tempc
  {#3}\ifx \@tempc \@empty \let \@tempc \@tempb \let \@tempb \@tempa \fi \ifx
  \@tempb \@empty \def\@tempb {arXiv}\fi \@ifundefined
  {mn@eprint@\@tempb}{\@tempb:\@tempc}{\expandafter \expandafter \csname
  mn@eprint@\@tempb\endcsname \expandafter{\@tempc}}}

\bibitem[\protect\citeauthoryear{{Alexander}}{{Alexander}}{2012}]{Alexander2012}
{Alexander} R.,  2012, \mn@doi [\apjl] {10.1088/2041-8205/757/2/L29}, \href
  {http://adsabs.harvard.edu/abs/2012ApJ...757L..29A} {757, L29}

\bibitem[\protect\citeauthoryear{{Anderson}, {Adams}  \& {Calvet}}{{Anderson}
  et~al.}{2013}]{Anderson2013}
{Anderson} K.~R.,  {Adams} F.~C.,   {Calvet} N.,  2013, \mn@doi [\apj]
  {10.1088/0004-637X/774/1/9}, \href
  {http://adsabs.harvard.edu/abs/2013ApJ...774....9A} {774, 9}

\bibitem[\protect\citeauthoryear{{Ansdell}, {Williams}, {Manara}, {Miotello},
  {Facchini}, {van der Marel}, {Testi}  \& {van Dishoeck}}{{Ansdell}
  et~al.}{2017}]{Ansdell2017}
{Ansdell} M.,  {Williams} J.~P.,  {Manara} C.~F.,  {Miotello} A.,  {Facchini}
  S.,  {van der Marel} N.,  {Testi} L.,   {van Dishoeck} E.~F.,  2017, \mn@doi
  [\aj] {10.3847/1538-3881/aa69c0}, \href
  {http://adsabs.harvard.edu/abs/2017AJ....153..240A} {153, 240}

\bibitem[\protect\citeauthoryear{{Armitage}, {Simon}  \& {Martin}}{{Armitage}
  et~al.}{2013}]{Armitage2013}
{Armitage} P.~J.,  {Simon} J.~B.,   {Martin} R.~G.,  2013, \mn@doi [\apjl]
  {10.1088/2041-8205/778/1/L14}, \href
  {http://adsabs.harvard.edu/abs/2013ApJ...778L..14A} {778, L14}

\bibitem[\protect\citeauthoryear{{Bai}}{{Bai}}{2016}]{Bai2016}
{Bai} X.-N.,  2016, \mn@doi [\apj] {10.3847/0004-637X/821/2/80}, \href
  {http://adsabs.harvard.edu/abs/2016ApJ...821...80B} {821, 80}

\bibitem[\protect\citeauthoryear{{Bai} \& {Stone}}{{Bai} \&
  {Stone}}{2013}]{Bai2013}
{Bai} X.-N.,  {Stone} J.~M.,  2013, \mn@doi [\apj]
  {10.1088/0004-637X/769/1/76}, \href
  {http://adsabs.harvard.edu/abs/2013ApJ...769...76B} {769, 76}

\bibitem[\protect\citeauthoryear{{Balbus} \& {Hawley}}{{Balbus} \&
  {Hawley}}{1991}]{Balbus1991}
{Balbus} S.~A.,  {Hawley} J.~F.,  1991, \mn@doi [\apj] {10.1086/170270}, \href
  {http://adsabs.harvard.edu/abs/1991ApJ...376..214B} {376, 214}

\bibitem[\protect\citeauthoryear{{Banzatti}, {Pascucci}, {Edwards}, {Fang},
  {Gorti}  \& {Flock}}{{Banzatti} et~al.}{2019}]{Banzatti2019}
{Banzatti} A.,  {Pascucci} I.,  {Edwards} S.,  {Fang} M.,  {Gorti} U.,
  {Flock} M.,  2019, \mn@doi [\apj] {10.3847/1538-4357/aaf1aa}, \href
  {http://adsabs.harvard.edu/abs/2019ApJ...870...76B} {870, 76}

\bibitem[\protect\citeauthoryear{{B{\'e}thune}, {Lesur}  \&
  {Ferreira}}{{B{\'e}thune} et~al.}{2017}]{Bet2017}
{B{\'e}thune} W.,  {Lesur} G.,   {Ferreira} J.,  2017, \mn@doi [\aap]
  {10.1051/0004-6361/201630056}, \href
  {http://adsabs.harvard.edu/abs/2017A%26A...600A..75B} {600, A75}

\bibitem[\protect\citeauthoryear{{Bitsch}, {Johansen}, {Lambrechts}  \&
  {Morbidelli}}{{Bitsch} et~al.}{2015}]{Bitsch2015}
{Bitsch} B.,  {Johansen} A.,  {Lambrechts} M.,   {Morbidelli} A.,  2015,
  \mn@doi [\aap] {10.1051/0004-6361/201424964}, \href
  {http://adsabs.harvard.edu/abs/2015A%26A...575A..28B} {575, A28}

\bibitem[\protect\citeauthoryear{{Blandford} \& {Payne}}{{Blandford} \&
  {Payne}}{1982}]{Blandford1982}
{Blandford} R.~D.,  {Payne} D.~G.,  1982, \mn@doi [\mnras]
  {10.1093/mnras/199.4.883}, \href
  {http://adsabs.harvard.edu/abs/1982MNRAS.199..883B} {199, 883}

\bibitem[\protect\citeauthoryear{{D'Alessio}, {Calvet}  \&
  {Hartmann}}{{D'Alessio} et~al.}{2001}]{D'Alessio2001}
{D'Alessio} P.,  {Calvet} N.,   {Hartmann} L.,  2001, \mn@doi [\apj]
  {10.1086/320655}, \href {http://adsabs.harvard.edu/abs/2001ApJ...553..321D}
  {553, 321}

\bibitem[\protect\citeauthoryear{{Fang} et~al.,}{{Fang}
  et~al.}{2018}]{Fang2018}
{Fang} M.,  et~al., 2018, \mn@doi [\apj] {10.3847/1538-4357/aae780}, \href
  {http://adsabs.harvard.edu/abs/2018ApJ...868...28F} {868, 28}

\bibitem[\protect\citeauthoryear{{Ferreira} \& {Pelletier}}{{Ferreira} \&
  {Pelletier}}{1995}]{Ferreira1995}
{Ferreira} J.,  {Pelletier} G.,  1995, \aap, \href
  {http://adsabs.harvard.edu/abs/1995A%26A...295..807F} {295, 807}

\bibitem[\protect\citeauthoryear{{Fromang}, {Latter}, {Lesur}  \&
  {Ogilvie}}{{Fromang} et~al.}{2013}]{Fromang2013}
{Fromang} S.,  {Latter} H.,  {Lesur} G.,   {Ogilvie} G.~I.,  2013, \mn@doi
  [\aap] {10.1051/0004-6361/201220016}, \href
  {http://adsabs.harvard.edu/abs/2013A%26A...552A..71F} {552, A71}

\bibitem[\protect\citeauthoryear{{Guilet} \& {Ogilvie}}{{Guilet} \&
  {Ogilvie}}{2014}]{Guilet2014}
{Guilet} J.,  {Ogilvie} G.~I.,  2014, \mn@doi [\mnras] {10.1093/mnras/stu532},
  \href {http://adsabs.harvard.edu/abs/2014MNRAS.441..852G} {441, 852}

\bibitem[\protect\citeauthoryear{{Hartmann} \& {Bae}}{{Hartmann} \&
  {Bae}}{2018}]{Hartmann2018}
{Hartmann} L.,  {Bae} J.,  2018, \mn@doi [\mnras] {10.1093/mnras/stx2775},
  \href {http://adsabs.harvard.edu/abs/2018MNRAS.474...88H} {474, 88}

\bibitem[\protect\citeauthoryear{{Hartmann}, {Calvet}, {Gullbring}  \&
  {D'Alessio}}{{Hartmann} et~al.}{1998}]{Hartmann1998}
{Hartmann} L.,  {Calvet} N.,  {Gullbring} E.,   {D'Alessio} P.,  1998, \mn@doi
  [\apj] {10.1086/305277}, \href
  {http://adsabs.harvard.edu/abs/1998ApJ...495..385H} {495, 385}

\bibitem[\protect\citeauthoryear{{Hasegawa}, {Okuzumi}, {Flock}  \&
  {Turner}}{{Hasegawa} et~al.}{2017}]{Hasegawa2017}
{Hasegawa} Y.,  {Okuzumi} S.,  {Flock} M.,   {Turner} N.~J.,  2017, \mn@doi
  [\apj] {10.3847/1538-4357/aa7d55}, \href
  {http://adsabs.harvard.edu/abs/2017ApJ...845...31H} {845, 31}

\bibitem[\protect\citeauthoryear{{Izidoro}, {Bitsch}, {Raymond}, {Johansen},
  {Morbidelli}, {Lambrechts}  \& {Jacobson}}{{Izidoro}
  et~al.}{2019}]{Izidoro2019}
{Izidoro} A.,  {Bitsch} B.,  {Raymond} S.~N.,  {Johansen} A.,  {Morbidelli} A.,
   {Lambrechts} M.,   {Jacobson} S.~A.,  2019, arXiv e-prints, \href
  {http://adsabs.harvard.edu/abs/2019arXiv190208772I} {}

\bibitem[\protect\citeauthoryear{{Khajenabi}, {Shadmehri}, {Pessah}  \&
  {Martin}}{{Khajenabi} et~al.}{2018}]{khajenabi2018}
{Khajenabi} F.,  {Shadmehri} M.,  {Pessah} M.~E.,   {Martin} R.~G.,  2018,
  \mn@doi [\mnras] {10.1093/mnras/sty153}, \href
  {http://adsabs.harvard.edu/abs/2018MNRAS.475.5059K} {475, 5059}

\bibitem[\protect\citeauthoryear{{Klahr} \& {Hubbard}}{{Klahr} \&
  {Hubbard}}{2014}]{Klahr2014}
{Klahr} H.,  {Hubbard} A.,  2014, \mn@doi [\apj] {10.1088/0004-637X/788/1/21},
  \href {http://adsabs.harvard.edu/abs/2014ApJ...788...21K} {788, 21}

\bibitem[\protect\citeauthoryear{{Konigl}}{{Konigl}}{1989}]{Konigl1989}
{Konigl} A.,  1989, \mn@doi [\apj] {10.1086/167585}, \href
  {http://adsabs.harvard.edu/abs/1989ApJ...342..208K} {342, 208}

\bibitem[\protect\citeauthoryear{{Kratter} \& {Lodato}}{{Kratter} \&
  {Lodato}}{2016}]{Krat2016}
{Kratter} K.,  {Lodato} G.,  2016, \mn@doi [\araa]
  {10.1146/annurev-astro-081915-023307}, \href
  {http://adsabs.harvard.edu/abs/2016ARA%26A..54..271K} {54, 271}

\bibitem[\protect\citeauthoryear{{Lesur}, {Ferreira}  \& {Ogilvie}}{{Lesur}
  et~al.}{2013}]{Lesur2013}
{Lesur} G.,  {Ferreira} J.,   {Ogilvie} G.~I.,  2013, \mn@doi [\aap]
  {10.1051/0004-6361/201220395}, \href
  {http://adsabs.harvard.edu/abs/2013A%26A...550A..61L} {550, A61}

\bibitem[\protect\citeauthoryear{{Lodato}, {Scardoni}, {Manara}  \&
  {Testi}}{{Lodato} et~al.}{2017}]{Lodato17}
{Lodato} G.,  {Scardoni} C.~E.,  {Manara} C.~F.,   {Testi} L.,  2017, \mn@doi
  [\mnras] {10.1093/mnras/stx2273}, \href
  {http://adsabs.harvard.edu/abs/2017MNRAS.472.4700L} {472, 4700}

\bibitem[\protect\citeauthoryear{{Lubow}, {Papaloizou}  \& {Pringle}}{{Lubow}
  et~al.}{1994}]{Lubow1994}
{Lubow} S.~H.,  {Papaloizou} J.~C.~B.,   {Pringle} J.~E.,  1994, \mn@doi
  [\mnras] {10.1093/mnras/267.2.235}, \href
  {http://adsabs.harvard.edu/abs/1994MNRAS.267..235L} {267, 235}

\bibitem[\protect\citeauthoryear{{Manara} et~al.,}{{Manara}
  et~al.}{2016}]{Manara2016}
{Manara} C.~F.,  et~al., 2016, \mn@doi [\aap] {10.1051/0004-6361/201628549},
  \href {http://adsabs.harvard.edu/abs/2016A%26A...591L...3M} {591, L3}

\bibitem[\protect\citeauthoryear{{Marcus}, {Pei}, {Jiang}, {Barranco},
  {Hassanzadeh}  \& {Lecoanet}}{{Marcus} et~al.}{2015}]{Marcus2015}
{Marcus} P.~S.,  {Pei} S.,  {Jiang} C.-H.,  {Barranco} J.~A.,  {Hassanzadeh}
  P.,   {Lecoanet} D.,  2015, \mn@doi [\apj] {10.1088/0004-637X/808/1/87},
  \href {http://adsabs.harvard.edu/abs/2015ApJ...808...87M} {808, 87}

\bibitem[\protect\citeauthoryear{{Mulders}, {Pascucci}, {Manara}, {Testi},
  {Herczeg}, {Henning}, {Mohanty}  \& {Lodato}}{{Mulders}
  et~al.}{2017}]{Mulders2017}
{Mulders} G.~D.,  {Pascucci} I.,  {Manara} C.~F.,  {Testi} L.,  {Herczeg}
  G.~J.,  {Henning} T.,  {Mohanty} S.,   {Lodato} G.,  2017, \mn@doi [\apj]
  {10.3847/1538-4357/aa8906}, \href
  {http://adsabs.harvard.edu/abs/2017ApJ...847...31M} {847, 31}

\bibitem[\protect\citeauthoryear{{Najita} \& {Bergin}}{{Najita} \&
  {Bergin}}{2018}]{Najita2018}
{Najita} J.~R.,  {Bergin} E.~A.,  2018, \mn@doi [\apj]
  {10.3847/1538-4357/aad80c}, \href
  {http://adsabs.harvard.edu/abs/2018ApJ...864..168N} {864, 168}

\bibitem[\protect\citeauthoryear{{Nelson}, {Gressel}  \& {Umurhan}}{{Nelson}
  et~al.}{2013}]{Nelson2013}
{Nelson} R.~P.,  {Gressel} O.,   {Umurhan} O.~M.,  2013, \mn@doi [\mnras]
  {10.1093/mnras/stt1475}, \href
  {http://adsabs.harvard.edu/abs/2013MNRAS.435.2610N} {435, 2610}

\bibitem[\protect\citeauthoryear{{Ogihara}, {Morbidelli}  \&
  {Guillot}}{{Ogihara} et~al.}{2015}]{Ogihara2015}
{Ogihara} M.,  {Morbidelli} A.,   {Guillot} T.,  2015, \mn@doi [\aap]
  {10.1051/0004-6361/201527117}, \href
  {http://adsabs.harvard.edu/abs/2015A%26A...584L...1O} {584, L1}

\bibitem[\protect\citeauthoryear{{Ogihara}, {Kokubo}, {Suzuki}  \&
  {Morbidelli}}{{Ogihara} et~al.}{2018}]{Ogihara2018}
{Ogihara} M.,  {Kokubo} E.,  {Suzuki} T.~K.,   {Morbidelli} A.,  2018, \mn@doi
  [\aap] {10.1051/0004-6361/201832720}, \href
  {http://adsabs.harvard.edu/abs/2018A%26A...615A..63O} {615, A63}

\bibitem[\protect\citeauthoryear{{Okuzumi}, {Takeuchi}  \& {Muto}}{{Okuzumi}
  et~al.}{2014}]{Okuzumi2014}
{Okuzumi} S.,  {Takeuchi} T.,   {Muto} T.,  2014, \mn@doi [\apj]
  {10.1088/0004-637X/785/2/127}, \href
  {http://adsabs.harvard.edu/abs/2014ApJ...785..127O} {785, 127}

\bibitem[\protect\citeauthoryear{{Okuzumi}, {Momose}, {Sirono}, {Kobayashi}  \&
  {Tanaka}}{{Okuzumi} et~al.}{2016}]{Okuzumi2016}
{Okuzumi} S.,  {Momose} M.,  {Sirono} S.-i.,  {Kobayashi} H.,   {Tanaka} H.,
  2016, \mn@doi [\apj] {10.3847/0004-637X/821/2/82}, \href
  {http://adsabs.harvard.edu/abs/2016ApJ...821...82O} {821, 82}

\bibitem[\protect\citeauthoryear{{Rafikov}}{{Rafikov}}{2017}]{Rafikov2017}
{Rafikov} R.~R.,  2017, \mn@doi [\apj] {10.3847/1538-4357/aa6249}, \href
  {http://adsabs.harvard.edu/abs/2017ApJ...837..163R} {837, 163}

\bibitem[\protect\citeauthoryear{{Rigliaco}, {Natta}, {Randich}, {Testi}  \&
  {Biazzo}}{{Rigliaco} et~al.}{2011}]{Rigliaco2011}
{Rigliaco} E.,  {Natta} A.,  {Randich} S.,  {Testi} L.,   {Biazzo} K.,  2011,
  \mn@doi [\aap] {10.1051/0004-6361/201015299}, \href
  {http://adsabs.harvard.edu/abs/2011A%26A...525A..47R} {525, A47}

\bibitem[\protect\citeauthoryear{{Shadmehri}, {Khajenabi}  \&
  {Pessah}}{{Shadmehri} et~al.}{2018a}]{shadmehri2018}
{Shadmehri} M.,  {Khajenabi} F.,   {Pessah} M.~E.,  2018a, \mn@doi [\apj]
  {10.3847/1538-4357/aad047}, \href
  {http://adsabs.harvard.edu/abs/2018ApJ...863...33S} {863, 33}

\bibitem[\protect\citeauthoryear{{Shadmehri}, {Ghoreyshi}  \&
  {Alipour}}{{Shadmehri} et~al.}{2018b}]{Mshadmehri2018}
{Shadmehri} M.,  {Ghoreyshi} S.~M.,   {Alipour} N.,  2018b, \mn@doi [\apj]
  {10.3847/1538-4357/aae2b5}, \href
  {http://adsabs.harvard.edu/abs/2018ApJ...867...41S} {867, 41}

\bibitem[\protect\citeauthoryear{{Shakura} \& {Sunyaev}}{{Shakura} \&
  {Sunyaev}}{1973}]{Shakura1973}
{Shakura} N.~I.,  {Sunyaev} R.~A.,  1973, \aap, \href
  {http://adsabs.harvard.edu/abs/1973A%26A....24..337S} {24, 337}

\bibitem[\protect\citeauthoryear{{Simon}, {Bai}, {Stone}, {Armitage}  \&
  {Beckwith}}{{Simon} et~al.}{2013a}]{Simon2013a}
{Simon} J.~B.,  {Bai} X.-N.,  {Stone} J.~M.,  {Armitage} P.~J.,   {Beckwith}
  K.,  2013a, \mn@doi [\apj] {10.1088/0004-637X/764/1/66}, \href
  {http://adsabs.harvard.edu/abs/2013ApJ...764...66S} {764, 66}

\bibitem[\protect\citeauthoryear{{Simon}, {Bai}, {Armitage}, {Stone}  \&
  {Beckwith}}{{Simon} et~al.}{2013b}]{Simon2013b}
{Simon} J.~B.,  {Bai} X.-N.,  {Armitage} P.~J.,  {Stone} J.~M.,   {Beckwith}
  K.,  2013b, \mn@doi [\apj] {10.1088/0004-637X/775/1/73}, \href
  {http://adsabs.harvard.edu/abs/2013ApJ...775...73S} {775, 73}

\bibitem[\protect\citeauthoryear{{Suzuki}, {Muto}  \& {Inutsuka}}{{Suzuki}
  et~al.}{2010}]{Suzuki2010}
{Suzuki} T.~K.,  {Muto} T.,   {Inutsuka} S.-i.,  2010, \mn@doi [\apj]
  {10.1088/0004-637X/718/2/1289}, \href
  {http://adsabs.harvard.edu/abs/2010ApJ...718.1289S} {718, 1289}

\bibitem[\protect\citeauthoryear{{Suzuki}, {Ogihara}, {Morbidelli}, {Crida}  \&
  {Guillot}}{{Suzuki} et~al.}{2016}]{Suzuki2016}
{Suzuki} T.~K.,  {Ogihara} M.,  {Morbidelli} A.,  {Crida} A.,   {Guillot} T.,
  2016, \mn@doi [\aap] {10.1051/0004-6361/201628955}, \href
  {http://adsabs.harvard.edu/abs/2016A%26A...596A..74S} {596, A74}

\bibitem[\protect\citeauthoryear{{Wang} \& {Goodman}}{{Wang} \&
  {Goodman}}{2017}]{Wang2017}
{Wang} L.,  {Goodman} J.~J.,  2017, \mn@doi [\apj]
  {10.3847/1538-4357/835/1/59}, \href
  {http://adsabs.harvard.edu/abs/2017ApJ...835...59W} {835, 59}

\makeatother
\end{thebibliography}


%%%%%%%%%%%%%%%%% APPENDICES %%%%%%%%%%%%%%%%%%%%%

\bsp	% typesetting comment
\label{lastpage}

\end{document}